\documentclass[12pt,preprint]{aastex}
\begin{document}

\baselineskip=0.8cm

\parindent=1.0cm

\title{THE COMPACT STAR-FORMING COMPLEX AT THE HEART OF NGC 253 
\altaffilmark{1,} \altaffilmark{2,} \altaffilmark{3}}

\author{T. J. Davidge}

\affil{Dominion Astrophysical Observatory,
\\National Research Council of Canada, 5071 West Saanich Road,
\\Victoria, BC Canada V9E 2E7\\tim.davidge@nrc.ca}

\altaffiltext{1}{Based on observations obtained at the Gemini Observatory, which is
operated by the Association of Universities for Research in Astronomy, Inc., under a
cooperative agreement with the NSF on behalf of the Gemini partnership: the National
Science Foundation (United States), the National Research Council (Canada), CONICYT
(Chile), the Australian Research Council (Australia), Minist\'{e}rio da Ci\^{e}ncia,
Tecnologia e Inova\c{c}\~{a}o (Brazil) and Ministerio de Ciencia, Tecnolog\'{i}a e
Innovaci\'{o}n Productiva (Argentina).}

\altaffiltext{2}{This research used the facilities of the Canadian Astronomy Data
Centre operated by the National Research Council of Canada with the support
of the Canadian Space Agency.}

\altaffiltext{3}{This research has made use of the NASA/IPAC Infrared Science Archive,
which is operated by the Jet Propulsion Laboratory, California Institute of Technology,
under contract with the National Aeronautics and Space Administration.}

\begin{abstract}

	We discuss integral field spectra of the compact star-forming complex 
that is the brightest near-infrared (NIR) source in the central regions of the 
starburst galaxy NGC 253. The spectra cover the $H$ and $K$ passbands
and were recorded with the Gemini NIR Spectrograph 
during sub-arcsec seeing conditions. Absorption features in the 
spectrum of the star-forming complex are weaker than in the surroundings. 
An absorption feature is found near $1.78\mu$m that coincides with 
the location of a C$_2$ bandhead. If this feature is due to C$_2$ then 
the star-forming complex has been in place for at least a 
few hundred Myr. Emission lines of Br$\gamma$, 
[FeII], and HeI$2.06\mu$m do not track the NIR continuum light. 
Pockets of star-forming activity that do not have associated concentrations of 
red supergiants, and so likely have ages $< 8$ Myr, are found along the western 
edge of the complex, and there is evidence that one such pocket contains a rich 
population of Wolf-Rayet stars. Unless the star-forming complex is significantly 
more metal-poor than the surroundings, then a significant fraction of its total 
mass is in stars with ages $< 8$ Myr. If the present-day star formation rate is 
maintained then the time scale to double its stellar mass ranges from a few Myr 
to a few tens of Myr, depending on the contribution made by stars older 
than $\sim 8$ Myr. If -- as suggested by some studies -- the star-forming complex is 
centered on the galaxy nucleus, which presumably contains a large population of 
old and intermediate-age stars, then the nucleus of NGC 253 is currently 
experiencing a phase of rapid growth in its stellar mass. 
\end{abstract}


\section{INTRODUCTION}

	Interactions and mergers can spur the large-scale re-distribution 
of material and angular momentum within a galaxy, thereby altering 
basic observational properties such as morphology. Some of the material that 
is displaced by such events will be channeled into the central regions of galaxies, 
where elevated levels of star-forming activity will be triggered in concert with 
those at large radii. Observations suggest that while elevated episodes of 
star-forming activity in a galaxy may die out first at large 
radii, it may continue for some time in the central regions 
(e.g. Soto \& Martin 2010). Given that the typical damping time 
for star burst activity in moderately low mass galaxies is many hundreds of Myr 
(McQuinn et al. 2010), then we might expect to find evidence for extensive 
star-forming activity in the central regions of late stage starbursts that is 
spread over a broad time period.

	The barred spiral galaxy NGC 253 is the dominant member of one of the three 
concentrations that collectively make up the Sculptor Group (Karachentsev et 
al. 2003). NGC 253 is experiencing a starburst (e.g. Rieke et al. 1980), and is 
one of the nearest such examples of this phenomenon. NGC 253 does not have an 
entourage of bright companions, and it is not clear what event (or events) triggered 
the starburst. Based on a survey of bright evolved stars in the NGC 253 disk, Davidge 
(2010) suggests that NGC 253 interacted with a now defunct companion within the past 
0.2 Gyr. Such an encounter could have spurred the formation of the 
bar, and caused the re-distribution of gas and dust throughout the galaxy
(e.g. Hopkins et al. 2009). Alternatively, a flyby encounter with another 
large galaxy could also have triggered bar formation and starburst 
activity (Lang et al. 2014). If a flyby happened more than $\sim 100$ 
Myr in the past then identifying the perturbing galaxy may be problematic. 

	The central regions of NGC 253 contain clues 
about the age and starburst triggering mechanism. 
The center of NGC 253 is heavily obscured at visible wavelengths, and so most 
previous studies have been conducted in the near-infrared 
(NIR). Engelbracht et al. (1998) used long-slit spectra and narrow-band images to 
investigate the ionization properties, stellar content and structural characteristics 
in a 7.5 arcsec radius aperture that contains most of the [NeII], Br$\gamma$, 
H$_2$, and mid-infrared (MIR) flux. They conclude that the interstellar medium (ISM) 
is excited by a radiation field that is powered by objects with T$_{eff} \leq 
38000$ K, indicating that a large population of young stars is present. 
This young component formed within the past $\sim 30$ Myr and contributes 
$\sim 10 - 20\%$ of the stellar mass in the region studied.

	There is a compact bright structure that defines the 
photometric center of NGC 253 in the NIR. This object, which has a sub-arcsec 
characteristic width, and its immediate surroundings are the subject of this paper. 
Following Gunthardt et al. (2015), we refer to this object as the IR Core 
(hereafter IRC). Modelling of Br$\gamma$ emission by Kornei \& McCrady 
(2009) suggests that the IRC contains a large number of ionizing stars that have an 
age of 5.7 Myr, and they estimate that its total stellar mass 
is $1.4 \times 10^7$M$_{\odot}$. Kornei \& McCrady (2009) 
further note that the IRC has been the site of an extended period of star-forming 
activity during recent epochs. This conclusion is based on the deep CO absorption 
bands that are seen in the integrated spectrum at $\lambda > 2.3\mu$m. 
Deep CO bands in the integrated spectra of young clusters 
originate in red supergiants (RSGs), and so are signatures of systems with ages 
older than $\sim 8 - 10$ Myr. 

	The IRC is offset by a few arcsec from what has been identified as the 
kinematic center of the galaxy (Muller-Sanchez et al. 2010; Rosenberg et al. 2013).
HOwever, Rosenberg et al. (2013) also find differences between the distributions of 
NIR continuum light, Br$\gamma$ emission, and H$_2$ emission, indicating that the 
stellar content near the center of NGC 253 has complex spatial and kinematic 
properties. Additional evidence of complexity is presented by Gunthardt et al. (2015), 
who find sub-structuring and a complicated projected distribution of stellar types 
in and around the IRC. They find that the IRC coincides with the rotational center 
of H$_2$ emission, and argue that the IRC is the true nucleus of NGC 253. 

	The complex projected distribution of stars and gas, coupled with the evidence 
for sub-structuring, demonstrates the need to map the distribution 
of spectroscopic features in and around the IRC. 
This task is well-suited to spectrographs that are equipped with 
an integral field unit (IFU). In the present study, deep IFU 
spectra of the central regions of NGC 253 that cover the $1.4 - 
2.3\mu$m wavelength interval are discussed. This wavelength interval contains 
absorption and emission features that can be used to trace the projected distribution 
of stars that span a wide range of ages, and identify 
areas of recent star-forming activity. 

	The nucleus of the late-type disk galaxy NGC 7793, which is the 
dominant member of another Sculptor Group concentration (Karachensheva et al. 2003), 
was also observed to represent the nucleus of a non-bursting, isolated spiral galaxy. 
To the best of our knowledge, these are the first NIR spectra of the central 
regions of NGC 7793. Bibby \& Crowther (2010) estimate from the H$\alpha$ flux 
that the SFR in NGC 7793 is 0.45 M$_{\odot}$ year$^{-1}$, and this 
is $\sim 6 \times$ lower than the SFR of NGC 253 measured by Ott et al. (2005). 
Carson et al. (2015) find that the photometric size of the NGC 7793 nucleus grows 
with decreasing wavelength, as expected if there is a spatially extended young 
population.

	Unlike NGC 253, there are indications that NGC 7793 has evolved 
in isolation throughout much of its life. The disk of NGC 7793 
spans many scale lengths (Vlajic et al. 2011), and Radburn-Smith et al. 
(2012) suggest that the outermost regions of its disk were populated by radial 
migration from smaller radii, rather than by the disruption of an earlier structure. 
Davidge (1998) found that the NGC 7793 disk contains 
stars that span a broad range of ages, and the oldest of these 
have an age $\sim 10$ Gyr. The age and radial extent of the NGC 7793 
disk are thus consistent with it having been in place for a long period of 
time and not having been disrupted by a merger or interaction. 

	Details of the observations and the reduction of the data 
can be found in Section 2. Archival NIR images that provide supplemental information 
for interpreting the spectra are discussed in Section 3. The strengths and on-sky 
distributions of spectroscopic features that are used to probe stellar content 
are discussed in Sections 4 and 5. Comparisons are made with model spectra in 
Section 6. A summary and discussion of the results are presented in Section 7.

\section{OBSERVATIONS \& REDUCTION}

	The spectra were recorded at Gemini South (GS) on the night of UT 
November 21, 2004 UT with the Gemini NIR Spectrograph (GNIRS; Elias et al. 2006) 
as part of program GS2004B-SV-22 (PI: Davidge). The GNIRS IFU is described in 
detail by Dubbeldam et al. (2000). The IFU was \footnote[1]{The IFU 
was destroyed during a run-away instrument warm-up in 2007.}
an image slicer design with 21 slitlets. It was used in 
low resolution mode for these observations, and so a $3.15 \times 4.85$ 
arcsec field was covered on the sky. Each slitlet is 0.15 arcsec wide and subtends 
32 pixels on the $1024 \times 1024$ InSb array in GNIRS, so that the sampling 
along the slit is 0.15 arcsec pixel$^{-1}$. Each spaxal thus 
subtends $0.15 \times 0.15$ arcsec.

	The IRC was positioned at the center of the IFU science field. 
While there are no point sources in the IFU observations from which the image 
quality can be measured directly, the IRC has a FWHM $\sim 0.6 - 0.7$ arcsec in 
the $K-$band continuum wavelength interval defined in Section 5.1. Given that the 
IRC is an extended object, then this sets an upper limit to the size of 
the seeing disk.

	A complete observation consisted of (1) five exposures of the target 
galaxy, with each exposure recorded at slightly different dither positions to assist 
with the identification and suppression of bad pixels and cosmic rays, followed by (2) 
five observations of a background sky field recorded with the same exposure times 
as the galaxy observations. This galaxy$+$sky sequence was repeated three times 
in each filter. Individual exposure times were 20 seconds for 
NGC 253 and 40 seconds for NGC 7793. The spectra presented here thus have 
total on-source integration times per filter of $5 \times 3 \times 20$ seconds $= 
300$ seconds for NGC 253, and $5 \times 3 \times 40$ seconds $=600$ seconds for 
NGC 7793.

	Ancillary data were recorded to monitor instrumental and atmospheric 
characteristics. An Ar lamp spectrum was recorded after each 
galaxy$+$sky pair, while a series of flat-field frames 
were recorded at various times during the night, with the slit illuminated by 
a light source in the Gemini Facility Calibration Unit. 
Telluric absorption features were monitored with spectra of the F4V 
star HR8999, which was observed at different times throughout the night. 
Hydrogen absorption lines in the HR8999 spectrum were removed 
by subtracting Voigt profiles that were fitted to those lines.

	The first step in the data reduction sequence was the 
removal of sky emission. A sky emission template was constructed 
for each galaxy spectrum by interpolating in time between the sky 
observations, and the result was subtracted from the spectrum. The 
sky-subtracted frames in a given filter at each dither position 
were then co-added, and the results were divided by a flat field 
frame to correct for pixel-to-pixel sensitivity variations and non-uniformities in 
optical throughput.

	The mean flat-fielded exposures were aligned by centering on the peak signal 
in each slitlet, and the results were combined by averaging the signal at 
each shifted pixel location. This was followed by wavelength calibration, which was 
done using a dispersion solution obtained from the arc spectra. The 
wavelength-calibrated galaxy spectra were then divided by the telluric star spectrum.
 
\section{NARROW-BAND IMAGING}

	The central regions of NGC 253 have a complex morphology, 
and even a single pointing with the GNIRS IFU samples a number of 
different structures. Moderately deep archival images of the center 
of NGC 253 were thus examined in an effort to better understand the region near 
the IRC. F128N and F164N WFC3 images that were recorded by the 
Hubble Space Telescope for program 12206 (PI: Westmoquette) 
and that had been processed with the WFC3 pipeline
were downloaded from the Canadian Astronomical Data Center \footnote[2]
{http://www.cadc-ccda.hia-iha.nrc-cnrc.gc.ca/en/hst/}. 
While a combination of broad and narrow-band images were recorded for this program, 
the central regions of NGC 253 are saturated in the broad-band images, 
and so only the narrow-band data are considered here.

	Three images were recorded in each filter, and these were aligned and 
median-combined on a filter-by-filter basis. The left hand panel of Figure 1
shows the central $6.7 \times 6.7$ arcsec of the stacked 
F164N image, with the IFU footprint indicated in black. 
The stacked F128N image was smoothed to match the 
angular resolution of the F164N image, and F128N-F164N colors are displayed 
in the right hand panel of Figure 1. 

\begin{figure}
\figurenum{1}
\epsscale{1.00}
\plotone{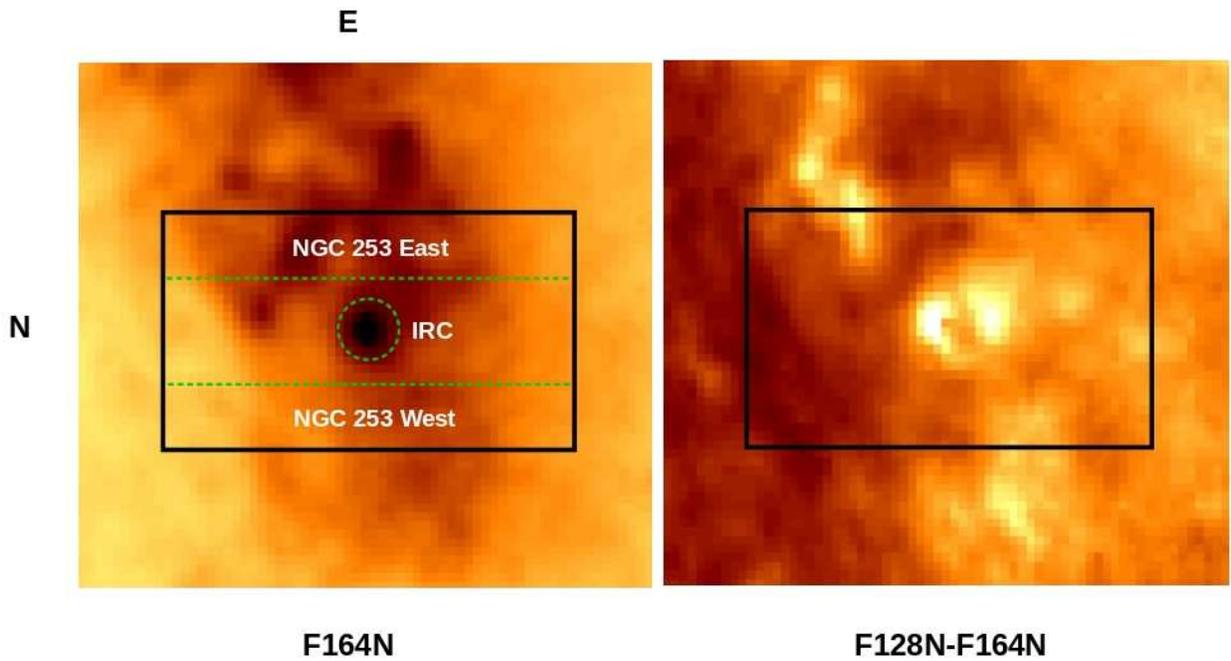}
\caption{Narrow-band HST WFC3 images of the central $6.7 \times 6.7$ arcsec 
of NGC 253. The left hand panel shows the F164N image with North to 
the left and East at the top. The area sampled by the IFU is outlined 
in black, while the areas from which the IRC, NGC 253 East, and NGC 253 
West spectra (Section 4) were extracted are shown in green. 
F128N--F164N color, displayed in magnitude units, is shown in the 
right hand panel, with redder colors having darker shading. There 
is a $\sim 0.9$ magnitude variation in F128N--F164N throughout the IFU field, and at 
least some of the structures in the F164N image are probably due to dust.}
\end{figure}

	The F128N--F164N color varies throughout the 
area sampled with the IFU. A filament with moderately red color runs E/W through 
the IRC, and is flanked by two pockets with blue F128N--F164N colors. The 
detection of areas with very blue colors in the IRC is perhaps not surprising given 
the spectroscopic evidence for very young stars there (Section 4). The red filament 
may be a dust lane, or a region between the blue pockets 
that is dominated by red stars. The filament has a color that is comparable to 
the area to the south of the IRC and is not as red as some of 
the suspected dust features, and this is consistent with the latter possibility.
That the F128N--F164N color varies in the IRC foreshadows one of the 
general results of this paper, which is that the age distribution of stars in 
and around the IRC varies with location -- the IRC is 
not a simple stellar population, but is instead a compact star-forming 
complex. Additional evidence to support this claim is presented in Section 5.

	Aside from the IRC, other less prominent structures are also sampled 
by the IFU, with most of these to the east and north 
of the IRC. Kornei \& McCrady (2009) note the presence of dust 
lanes within a few arcsec of the IRC in their NIR images, and it is likely that 
some of the structures in Figure 1 are not physically distinct entities, but 
may instead be artifacts of dust absorption. The regions between the IRC and the 
knots that are $\sim 1$ arcsec to the north east of the IRC are separated 
by a filament that -- owing to its very red color -- is a candidate dust lane. 
The north east corner of the IFU field contains an area with a 
blue F128N--F164N color, and in Section 5 it is demonstrated that this area 
shows spectroscopic signatures of very recent star formation. Finally, the north 
west corner of the IFU samples a region with red F128N--F164N 
colors, and a filamentary structure that appears to be a 
dust lane is also seen there in the F164N image. 

\section{RESULTS: EXTRACTED SPECTRA}

	Using the F164N image in Figure 1 as a guide, spectra 
that sample three different regions were extracted 
for detailed investigation: (1) the IRC, (2) the eastern portion of the IFU, where 
faint sub-structures are evident, and (3) the western portion of the IFU, where 
the light distribution is more uniform than elsewhere in the area sampled. 
The IRC spectrum was constructed by summing signal within a 0.4 arcsec radius 
circular gather centered on the IRC, which is 
where the signal was at least one half of the peak value. 
The other two spectra were obtained by combining 
signal in the $0.8 \times 4.85$ arcsec areas at the eastern and western edges of the 
IFU. These regions thus contain signal from sources that are at least 0.8 
arcsec from the center of the IRC, and so contamination from 
the IRC should not be a concern. The extraction regions 
are indicated in Figure 1. The spatial distributions of prominent 
spectroscopic features at finer angular resolutions are examined in Section 5, 
while the extracted spectra are compared with model spectra in Section 6.

\subsection{The IRC and the Nucleus of NGC 7793}

	The extracted spectra of the IRC are shown in Figures 2 ($H$) and 3 
($K$), where prominent emission and absorption features are identified. 
The continuum has been divided out using a low-order function that was 
fit to each spectrum. A high-order continuum function was not 
required as high frequency components in the system response 
were removed during the division by the telluric standard. Spectra of the NGC 7793 
nucleus are shown at the bottom of the figures. The NGC 7793 spectra were
extracted by summing the signal in a 0.4 arcsec radius aperture, 
which is where the signal is at least one half of the peak value.

\begin{figure}
\figurenum{2}
\epsscale{1.00}
\plotone{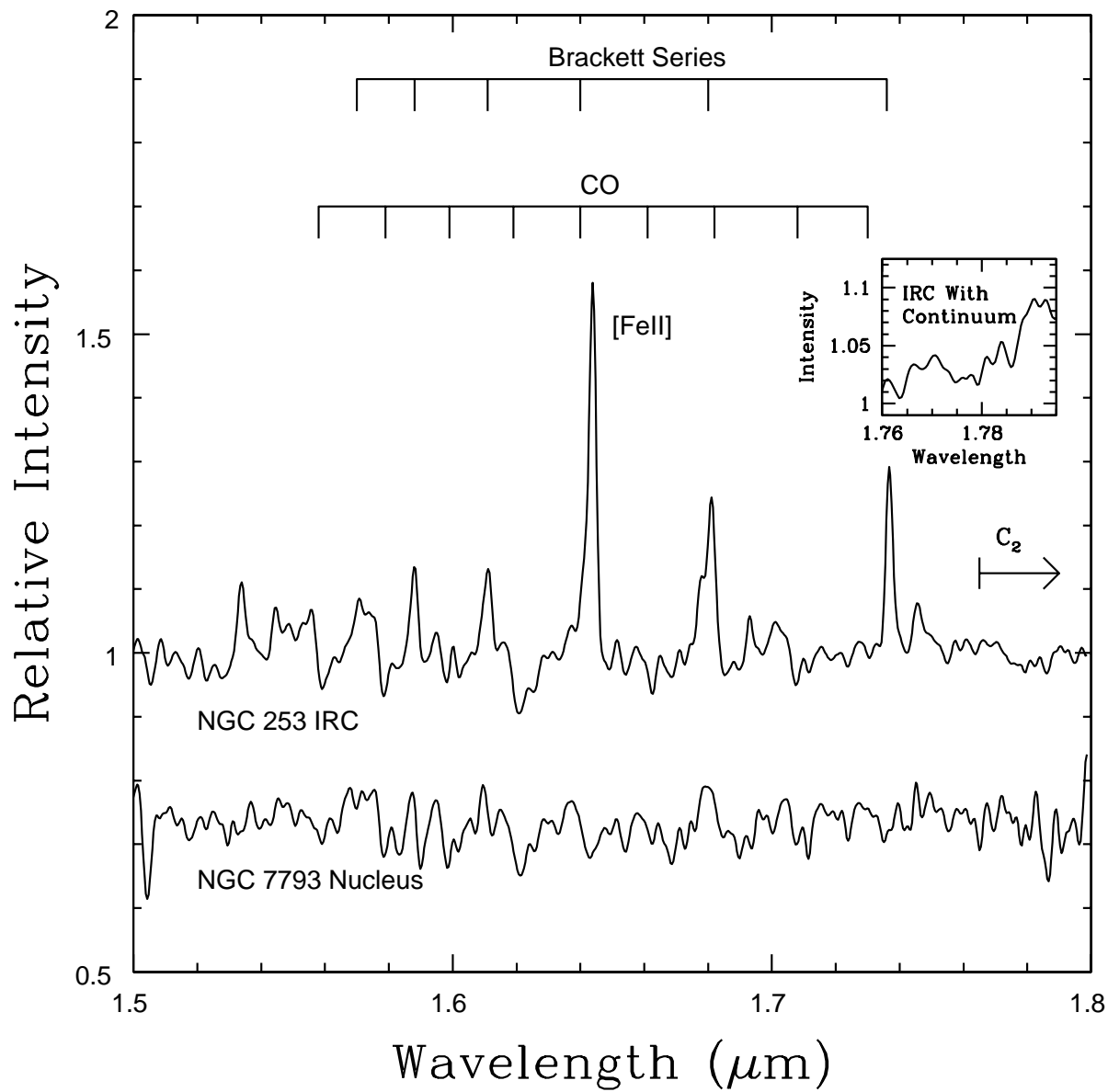}
\caption{$H-$band spectra of the IRC and the nucleus of NGC 7793. The IRC 
spectrum was obtained by combining signal in a 0.4 arcsec radius circular gather 
that is centered on the IRC. The spectra have been shifted along the vertical 
axis for display purposes. The inset shows the IRC spectrum before continuum removal
in the wavelength interval where C$_2$ absorption is expected. The absorption trough 
that is evident in the continuum-corrected spectrum is clearly seen in the inset, 
indicating that this feature is not an artifact of continuum removal.}
\end{figure}

\begin{figure}
\figurenum{3}
\epsscale{1.00}
\plotone{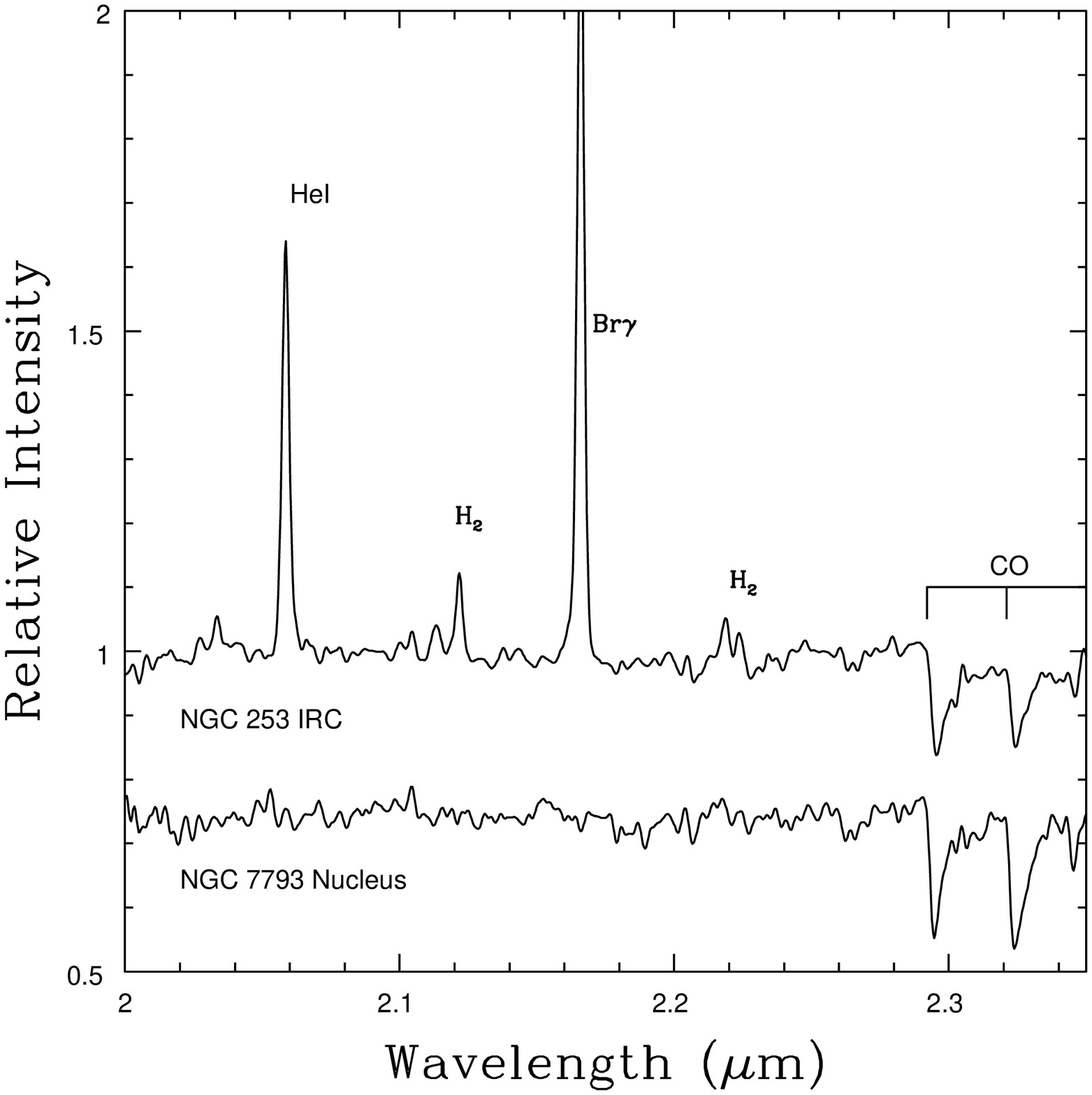}
\caption{Same as Figure 2, but showing $K-$band spectra. Note that the CO bands 
in the IRC spectrum are not as deep as in the NGC 7793 spectrum.}
\end{figure}

	There are obvious differences between the IRC and NGC 7793 spectra. 
Whereas there is a rich mix of absorption and emission features in the IRC spectra, 
there is no obvious line emission in the NGC 7793 spectra. 
This suggests that the nucleus of NGC 7793 does not contain a large 
population of stars with ages $< 10$ Myr. The first and second overtone bands of 
CO, which are the most prominent absorption features in this wavelength region, 
are deeper in NGC 7793 than in the IRC. However, a direct comparison of absorption 
features is affected by nebular continuum emission, 
which causes absorption features in the IRC spectrum to appear 
weaker than they would be in the absence of this emission. The 
contribution made by continuum emission to the spectra can be estimated 
in a purely empirical manner using the line and continuum measurements made 
by Armand et al. (1996), who observed five compact Galactic HII regions. The compact 
morphology of these HII regions makes them areas of concentrated 
line emission, and so at least some of the conditions near the 
center of NGC 253 may be replicated, albeit at much smaller levels of activity.

	The relative strengths of emission lines and 
the continuum depends on environmental properties such as 
the temperatures of the ionizing stars and the density of gas and dust. 
One way to identify a reference HII region for comparison with the IRC would be 
to consider the ultraviolet spectral-energy distribution (SED), as this is the 
wavelength range where the photons that power the emission originate. However, such 
a comparison is not possible for NGC 253 given the heavy obscuration towards the 
center of that galaxy. Photometric measurements made in 
the MIR, where the signal is dominated by thermal emission from 
dust that is heated by hot stars, offer an alternative. Both the hardness of the 
radiation field and the properties of the dust affect the MIR emission, 
and a template HII region was selected based on the W1 -- W2 color obtained 
from WISE (Wright et al. 2010) All Sky Survey images.

	The W1--W2 color of the HII region IRAS 12073-6233 matches the 
W1--W2 color of the center of NGC 253 to within 0.06 magnitudes. In contrast, 
IRAS 17200-3550 and 17455-2800 have W1--W2 colors that are 0.3 -- 0.4 magnitudes 
redder than the center of NGC 253. Thus, IRAS 12073--6233 was selected to 
serve as the template for estimating the contribution made by nebular continuum 
emission to the NIR light from the IRC.

	The Br$\gamma$ line width in the extracted IRC spectrum is $0.02\mu$m, 
and so -- using the entries for IRAS 12073-6233 in Table 1 and 2 of Armand 
et al. (1996) -- the nebular continuum is $\sim 6\%$ of the mean strength of the 
Br$\gamma$ line flux. The nebular continuum then accounts for $\sim 10\%$ of 
the signal from the IRC in $K$. The Starburst99 models summarized in Figure 8 
of Reines et al. (2010) predict that this level of continuum emission near $2\mu$m 
is expected for a system with an age 6 -- 8 Myr. 

	Rather than identifying a single HII region to serve as a reference object, an 
alternate approach might be to combine the results for all of the HII regions in the 
Armand et al. (1996) sample to obtain a mean line-to-continuum emission ratio. 
In fact, the ratio of the Br$\gamma$ to continuum strength in IRAS 12073-6233 
is close to the mean value of all 5 HII regions. Hence, adopting mean values 
from all HII regions would yield similar results to those obtained using 
IRAS 12073-6233 alone. 

	A 10\% contribution from the nebular continuum has a non-trivial effect on 
the absorption spectrum. This is demonstrated in Figures 4 and 5, where a featureless 
continuum that accounts for 10\% of the total light has been added to the NGC 7793 
spectra. Not surprisingly, the addition of this continuum weakens the absorption 
features in NGC 7793, producing a better match with the IRC spectra. 

\begin{figure}
\figurenum{4}
\epsscale{1.00}
\plotone{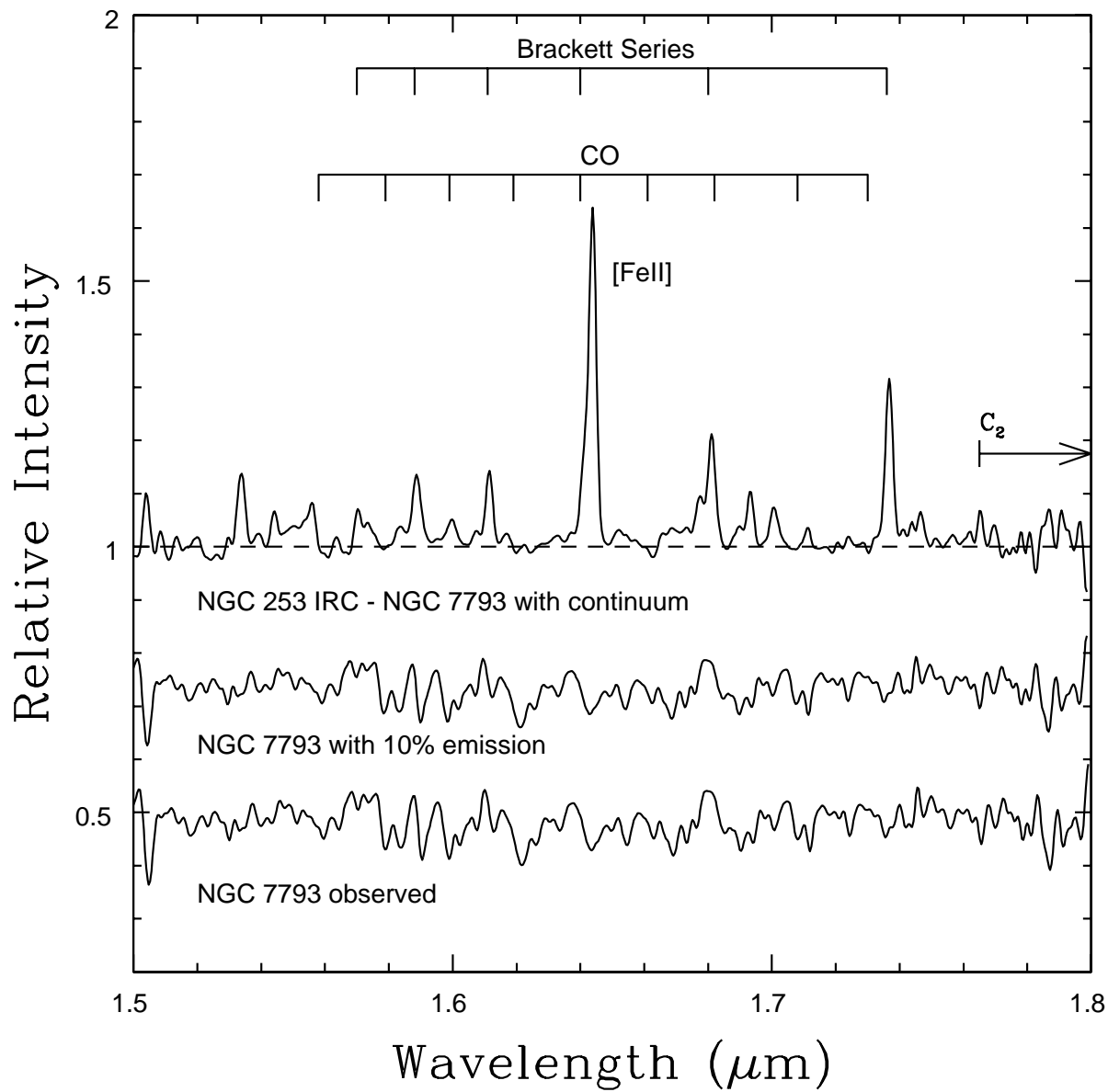}
\caption{Comparing absorption and emission features in the NGC 7793 and NGC 253 
IRC $H-$band spectra. The observed spectrum of the NGC 7793 nucleus is shown at the 
bottom. The middle spectrum shows the result of adding a featureless continuum that 
accounts for 10\% of the light to the NGC 7793 spectrum. The top spectrum is the 
difference between the IRC and the NGC 7793 $+$ continuum spectra; the 
dashed line indicates zero signal. The absorption features in the 
NGC 7793$+$continuum spectrum appear to match the depths of 
those in the NGC 253 IRC spectrum.}
\end{figure}

\begin{figure}
\figurenum{5}
\epsscale{1.00}
\plotone{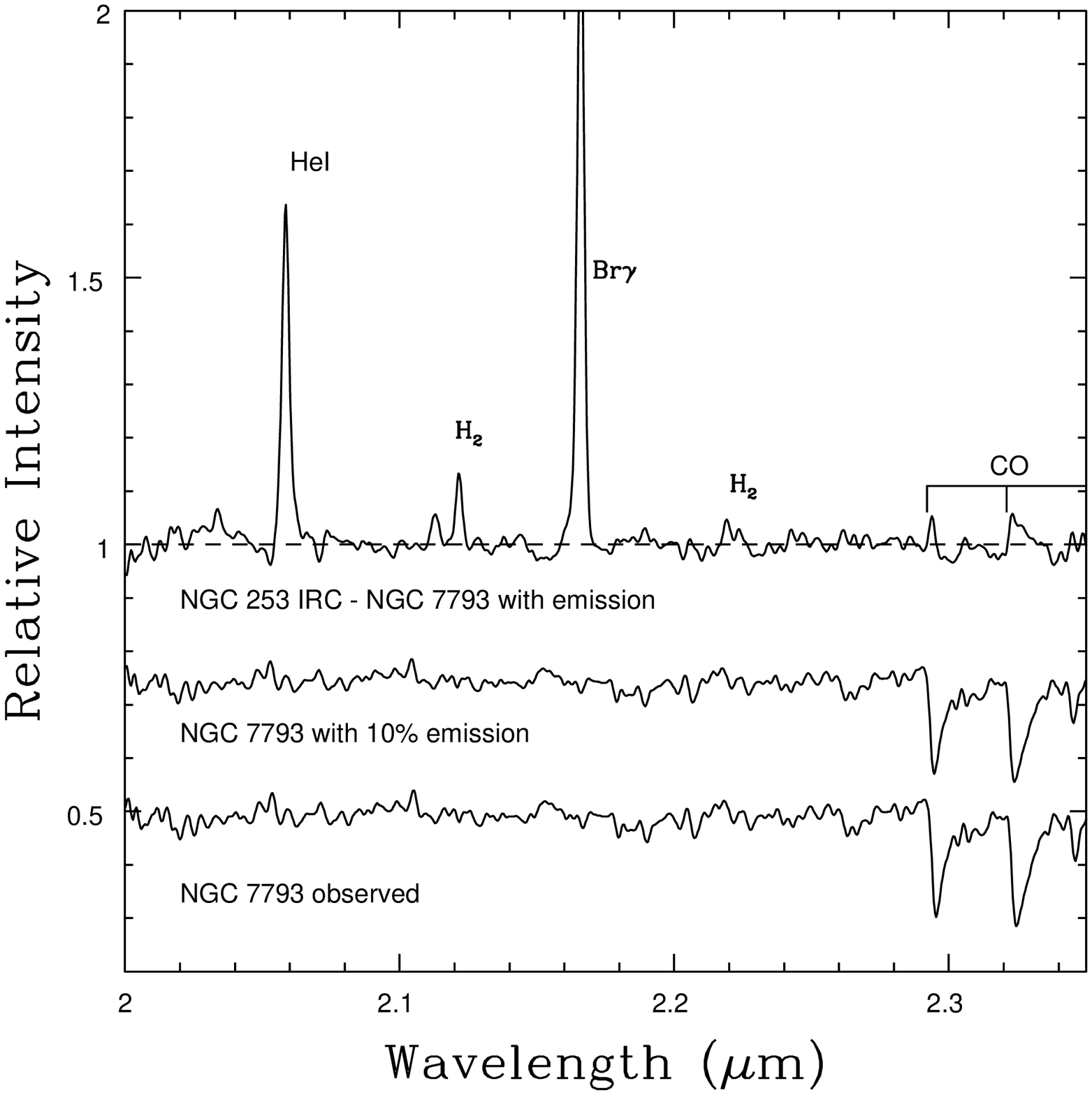}
\caption{Same as Figure 4, but showing $K$-band spectra. The tendency for 
the residuals to increase towards longer wavelengths when $\lambda > 2.3\mu$m 
may be a consequence of emission from hot dust near the center of NGC 253.}
\end{figure}

	The agreement between the absorption features 
in the IRC and the NGC 7793 spectra with a 10\% 
contribution from nebular continuum is examined in Figures 4 
and 5, where the differences between the IRC 
and the NGC 7793 nucleus $+$ continuum spectra are shown. Subtracting the 
NGC7793 $+$ continuum spectrum from the IRC spectrum results in 
a reasonable suppression of absorption features in the $H-$band. 
Weak emission features in the NGC 253 spectrum can also be more cleanly 
identified when the absorption spectrum is suppressed.

	The absorption features in the $K-$band are not as well matched as those 
in the $H-$band. In particular, the residuals near the 2.32$\mu$m (3--1) CO band 
indicate that this feature is stronger in the NGC 7793 spectrum than in the 
IRC spectrum, although the residuals near the 2.295$\mu$m (2--0) band indicate 
better agreement. A wavelength-dependent trend in the residuals could result if 
there is a significant thermal component in the IRC continuum. The experiments 
considered here assume a nebular component that contributes a fixed percentage of 
the light. In fact, there is significant thermal emission associated 
with star-forming knots in the central regions of NGC 253 
(e.g. Fernandez-Ontiveros et al. 2009), and this emission may 
contribute significantly near the long wavelength end of the $K-$band. 
The residuals near 2.32$\mu$m in Figure 5 might then 
be the consequence of failing to account for thermal emission. 

	The comparisons in Figures 4 and 5 ostensibly suggest that the brightest 
red stars in the IRC have spectroscopic properties that 
more-or-less match those in the nucleus of NGC 7793. This agreement is 
somewhat surprising, as it suggests similarity in the red stellar content. 
More specifically, if the NIR light in both systems is dominated by red stars then 
the agreement in line strengths could indicate comparable metallicities. However, 
the total $K$ magnitudes of these galaxies differ by $\sim 3$ magnitudes, in the 
sense that NGC 253 is the brighter of the two (Jarrett et al. 2003), and so the 
IRC might be expected to have a higher metallicity than the nucleus of NGC 7793. 
Perhaps the gas from which stars in the IRC are forming is 
moderately metal-poor, possibly having originated in an accreted companion. 
However, a problem with this speculation is that there 
is no evidence for metal-poor gas outside of the IRC. Indeed, 
the CO bands in the area around the IRC are deeper than those in the 
nucleus of NGC 7793 (Section 4.2), and have depths that are 
consistent with a super-solar metallicity (Section 6).

	A basic assumption in the preceeding paragraph is that the NIR 
continuum in the IRC is dominated by red stars. However, if the contribution from 
hot stars to the total NIR light is sufficiently large then absorption features 
from red stars will be veiled by the nearly featureless stellar continuum 
associated with OB stars. This can occur if there are large numbers of stars 
with ages $< 8$ Myr, as RSGs do not form in this age 
range -- the composite SED of such a young population is then 
dominated by hot stars. In fact, it is evident from the right hand panel of 
Figure 1 that the IRC has relatively blue NIR colors. The effect 
of a contribution from a large population of hot stars to the NIR spectrum of 
the IRC is investigated further in Section 6.

	The Ballick-Ramsey C$_2$ bandhead near $1.78\mu$m is 
a prominent absorption feature in C star spectra, and so might be expected in the 
spectrum of a system that experienced large-scale star formation during intermediate 
epochs. In fact, an absorption trough with a depth of a few percent is seen 
in the IRC spectrum near 1.78$\mu$m. Absorption at this wavelength is also seen 
in the East and West spectra (Section 4.2).

	The absorption trough near 1.78$\mu$m is not 
an artifact of continuum removal. This is demonstrated in the 
Figure 2 inset, where the IRC spectrum between 1.76 and 1.79$\mu$m 
without continuum removal is shown. The absorption trough near $1.78\mu$m 
is clearly seen in the spectrum with the continuum in place. Absorption near 
1.78$\mu$m is also evident in the $H-$band IRC spectrum in 
Figure 2 of Kornei \& McCrady (2009).

	The spectra of K and M giants do not show obvious absorption features at 
this wavelength (Rayner et al. 2009). However, it is still not obvious that C$_2$ 
has been detected in the IRC, given the very high levels of extinction 
towards the center of NGC 253. Engelbrecht et al. (1998) estimate that A$_V = 
9.2 \pm 0.7$ within 7.5 arcsec of the nucleus based on the Brackett decrement, 
and A$_V = 17 - 19$ from NIR colors. Kornei \& McCrady (2009) 
estimate that A$_V = 18 \pm 3$ from [FeII] lines, and A$_V \sim 12$ from NIR 
colors. Because of the substantial obscuration, there will likely be significant 
absorption by diffuse interstellar bands (DIBs). Geballe et al. (2011) discuss the 
detection of DIBs between 1.77 and $1.79\mu$m in the spectra of Galactic stars with 
A$_V \sim 20$ magnitudes, and an absorption feature at $1.793\mu$m is the deepest 
DIB discovered in their spectrum of GCS3-2. 

	Given that the extinction towards the center of NGC 253 is 
comparable to that towards the stars examined by Geballe et al. (2011) then 
the C$_2$ wavelength region likely also contains 
DIB signatures. However, these signatures are not expected to be present 
in the nuclear spectrum of NGC 7793, where the line-of-sight extinction is 
much lower than towards the IRC. The residuals in 
Figure 4 suggest that the IRC and the NGC 7793 nucleus 
spectra have similar levels of absorption in the wavelength interval that hosts 
the C$_2$ band.

\subsection{The Circumcluster Spectrum}

	The spectra of NGC 253 East and West are shown in Figures 6 ($H-$band) 
and 7 ($K-$band). The difference between the two spectra is shown 
at the bottom of each figure. The largest differences between the spectra 
occur at wavelengths where there are strong emission lines, with many (but 
not all) of these lines being stronger in NGC 253 East, which is an area where bluer 
F128N--F164N colors are seen (Figure 1). As for absorption features, the CO bands 
in the East and West spectra have similar depths. 
The scatter in the $1.76 - 1.78\mu$m interval, which covers part 
of the wavelength range where C$_2$ absorption might be expected, is similar 
to that seen at other wavelengths. However, there is 
much larger scatter in the differenced spectra at wavelengths $> 1.78\mu$m. 
This likely is due to the amplification of the noise that occurs near $1.8\mu$m 
when correcting for telluric water absorption. 

\begin{figure}
\figurenum{6}
\epsscale{1.00}
\plotone{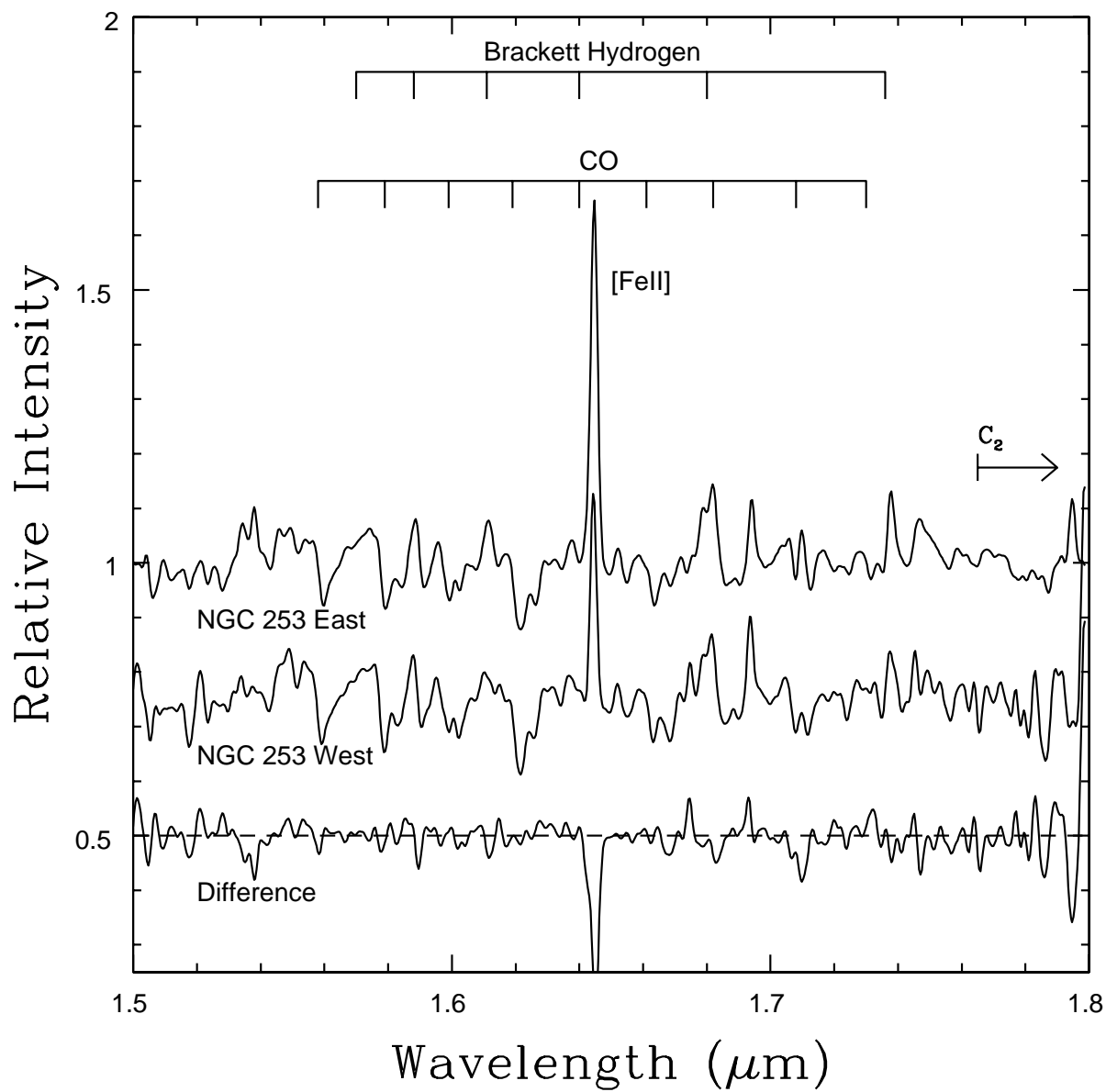}
\caption{Spectra of NGC253 East and West. The continuum has been divided out of 
each spectrum, and the difference between the two spectra is shown at the bottom of 
the figure. By far the largest differences are associated with [FeII], with this 
line being stronger in NGC 253 East. In contrast, the second-overtone CO bands in 
the East and West spectra have similar depths, and these are in turn deeper than in 
the IRC.}
\end{figure}

\begin{figure}
\figurenum{7}
\epsscale{1.00}
\plotone{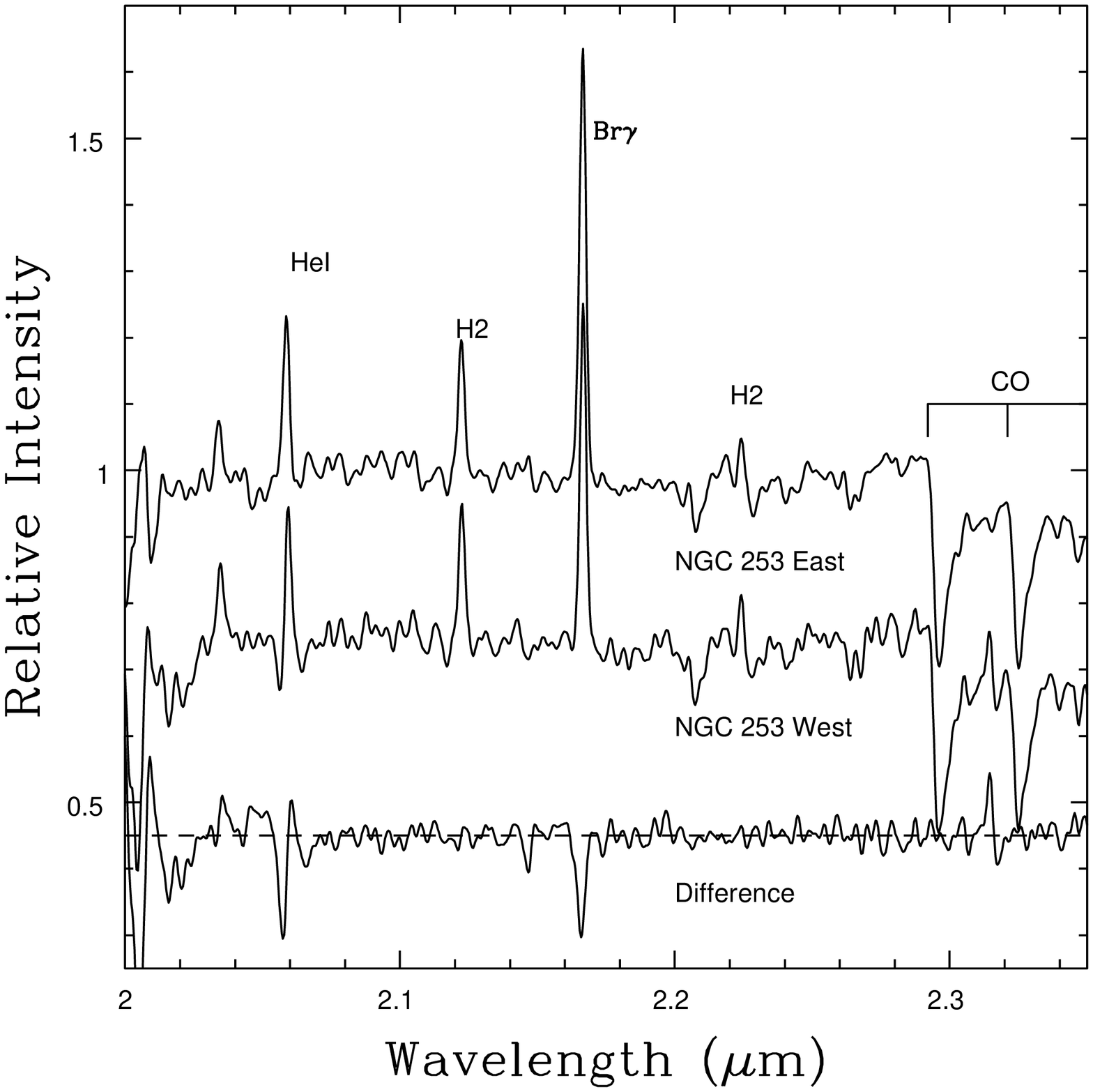}
\caption{Same as Figure 6, but showing $K-$band spectra. The East and West spectra 
have significantly different Br$\gamma$ and HeI strengths, with both lines being 
stronger in NGC 253 East. This is in contrast to the H$_2$ lines, which have 
remarkably similar strengths in NGC 253 East and West. 
The Br$\gamma$ and HeI features in both spectra are weaker than in the IRC. 
The first overtone CO bands in NGC 253 East and West have similar 
depths, and these are in turn much deeper than in the IRC.}
\end{figure}

	Many absorption and emission features in the $H$ and $K$ spectra of NGC 253 
East and West have markedly different strengths when compared with 
their counterparts in the IRC spectra. Emission lines of H, HeI, and [FeII] in NGC 253 
East and West are weaker than in the IRC. This suggests that 
while very young stars are present outside of the IRC, their density 
when measured with respect to the stars that dominate the NIR continuum is lower 
than in the IRC. H$_2$ emission is an exception. 
The strongest H$_2$ features in NGC 253 East and West 
are comparable to or even stronger than those in the IRC, with no 
difference between NGC 253 East and West. This suggests that the source 
of the H$_2$ excitation across the IFU field is well-mixed with the stars that 
dominate the NIR light when averaged over arcsec angular scales. As noted by 
Rosenberg et al. (2013), this is consistent with the stars that power H$_2$ emission 
having lower masses than those that power Br$\gamma$ and HeI emission, as these 
objects would then have had more time to mix throughout the central regions of NGC 253.

	The CO bands in NGC 253 East and West are deeper 
than in the IRC, and this is most noticeable for the 
first-overtone transitions. It is unlikely that this difference is 
a consequence of veiling of the IRC spectrum by non-stellar continua, 
which was discussed in Section 4.1. In order for nebular continuum emission 
alone to veil the CO bands in NGC 253 East and West so that they match what is 
seen in the IRC then the contribution from this emission would have to be $\sim 
7\times$ higher than estimated for the IRC in Section 4.1. 
Thermal emission also can not explain the differences in the depths of 
the $2.2\mu$m features, as the second overtone bands in the $H-$band circumnuclear 
spectra are deeper than in the IRC. The impact of a thermal contribution 
at these shorter wavelengths is negligible.
Rather than a non-stellar continuum, the differences between the IRC and 
East and West spectra could be due to large numbers of hot stars in the IRC, and 
this is considered in Section 6.

\section{THE SPATIAL DISTRIBUTION OF SPECTRAL FEATURES}

	The spectra that were investigated in Section 4 were constructed 
by combining data over large portions of the area sampled by the IFU. This 
binning was done at the expense of angular resolution to boost the S/N ratio. 
However, there are some features (most of them emission lines) that 
are sufficiently strong that their projected distribution can be examined 
without binning. The on-sky behaviour of these features is examined in this section.

\subsection{Continuum and Absorption Features}

	There are wavelength intervals that are free of strong emission 
features, and the signal in these can be used to map the distribution of light 
from the stars that dominate the NIR continuum free of contamination from 
prominent emission lines. Two such wavelength intervals are between $1.72\mu$m 
and $1.73\mu$m (hereafter referred to as Hcont) and $2.24\mu$m and $2.28\mu$m 
(hereafter Kcont). The distribution of light in Kcont and the Hcont--Kcont 
color are shown in Figure 8. The Kcont and Hcont--Kcont images show features that 
match those in Figure 1. The Hcont--Kcont color of the IRC is $0.3 - 0.4$ 
magnitudes bluer than that of the surroundings.

\begin{figure}
\figurenum{8}
\epsscale{0.50}
\plotone{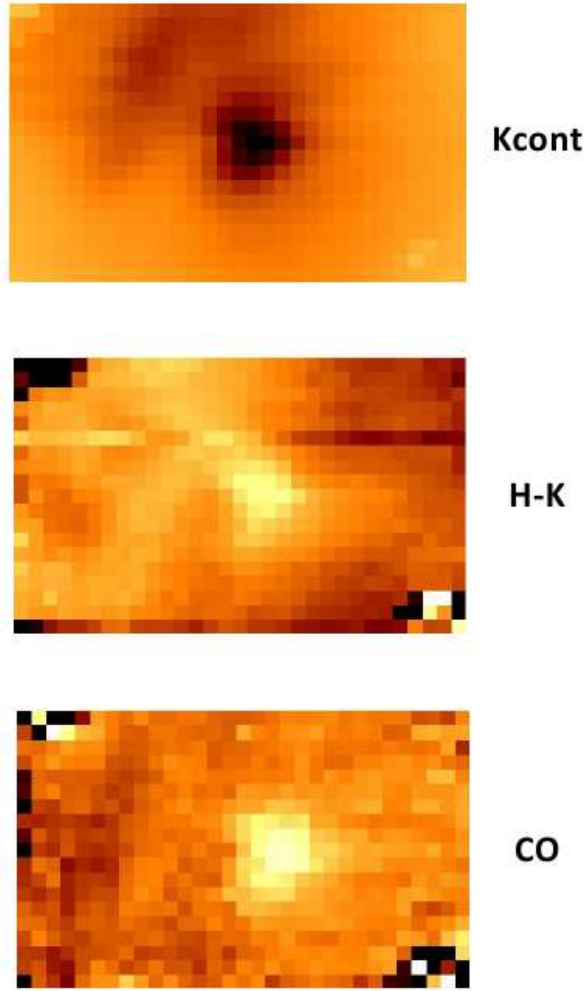}
\caption{Comparing the distribution of stellar continuum light and the depth of the 
(2,0) CO band. Each image covers $4.9 \times 3.2$ arcsec, with North to the left 
and East at the top. Darker shading in the Hcont--Kcont image indicates redder colors. 
Hcont--Kcont varies throughout the area surveyed, likely due to 
non-uniform extinction and variations in stellar content. 
The IRC is $\sim 0.3 - 0.4$ magnitudes bluer than the surroundings. 
The CO index, which gauges the strength of the 
(2,0) CO band head near $2.3\mu$m, is shown in 
the bottom panel. Smaller CO indices (i.e. shallower CO 
bands) are displayed with lighter colors. The CO index is weakest in the 
IRC, and is strongest near the northern boundary of the IFU. The CO index of the 
IRC is 10 -- 15\% lower than in the surroundings.}
\end{figure}

	Mapping the distribution of absorption features provides important 
information about stellar content and the mix of stellar types. 
However, obtaining maps for all but the deepest absorption features is 
problematic given the modest S/N ratio in spaxals that are external to the IRC. 
The first overtone CO bands, which are the 
deepest absorption features in these spectra, are an exception. The depth of 
the (2,0) CO band was measured in the wavelength interval $2.29\mu$m 
-- $2.31\mu$m, and a CO index was constructed by dividing the mean signal in this 
wavelength interval by the Kcont flux. The spatial behaviour of the CO index near 
the center of NGC 253 is shown in the lower panel of Figure 8.

	The spectroscopic properties of the IRC differ 
from those in the surrounding area. The CO index in the IRC is 10 -- 15\% 
weaker than in the surroundings, in agreement with the extracted spectra discussed in 
Sections 4.1 and 4.2. It is perhaps surprising that there 
are only minor variations in the CO index in the area external to 
the IRC given the large range of F128N--F164N colors that are seen there in Figure 1. 
An exception is the pocket of deep CO absorption near 
the northern edge of the IFU, which has a slightly 
redder Hcont--Kcont color than its surroundings. If there are large numbers of 
RSGs in this area that have ages $\sim 10$ Myr, and so are younger than those 
that populate most of the IFU field (Section 6), then this area might be 
expected to have deeper CO bands and redder colors.

	Kcont, Hcont--Kcont, and CO images were also constructed for the central 
regions of NGC 7793. Apart from an obvious nucleus in the Kcont image, 
these images do not reveal significant structure, and so are not shown here. 
In particular, the CO index of the NGC 7793 nucleus is comparable to that in the 
circumnuclear area, which is contrary to what is seen in NGC 253.

\subsection{Line emission}

	The emission lines in the NGC 253 spectra probe a diverse 
range of environments, and so may not track light 
from the stars that dominate the NIR continuum. In this 
section various emission line indices normalized to Kcont are used to examine 
differences in the distribution of the line-emitting sources and the underlying 
stellar continuum. Comparisons of this nature provide insights into 
the mix of stellar types throughout the area sampled with the IFU, including the 
identification of areas that have very young luminosity-weighted ages. 

	Spectral indices were computed for four emission lines using the 
wavelength intervals listed in Table 1. The line measurements were made in 
the wavelength interval between the points where the wings of the line merges 
with the adjacent continuum in these data. The continuum is measured 
on both sides of each line, and the mean continuum 
contribution was subtracted to obtain an index that tracks only line emission. 

\begin{table*}
\begin{center}
\begin{tabular}{lccc}
\tableline\tableline
Feature & Line & Continuum 1 & Continuum 2 \\
 & ($\mu$m) & ($\mu$m) & ($\mu$m) \\
\tableline
Br13 & 1.6136--1.6449 & 1.6088--1.6131 & 1.6454--1.6499 \\
$[$FeII$]$ & 1.6413--1.6485 & 1.6374--1.6408 & 1.6490--1.6524 \\ 
HeI & 2.0553--2.0637 & 2.0508--2.0547 & 2.0643--2.0675 \\
Br$\gamma$ & 2.1612--2.1705 & 2.1429--2.1606 & 2.1711--2.1796 \\
\tableline
\end{tabular}
\caption{Wavelengths for Emission Line Measurements}
\end{center}
\end{table*}

\subsubsection{Br$\gamma$/Kcont}

	The ratio of the Br$\gamma$ and Kcont indices 
is shown in the upper left hand corner of Figure 9. If Br$\gamma$ and Kcont 
have the same projected distributions then the image would be featureless. However, 
the Br$\gamma$ to Kcont ratio peaks in the outer areas of the IRC, 
with the highest ratios defining an arc that girds the western half of the IRC.
Br$\gamma$ emission thus has a more extended spatial 
distribution than Kcont in and around the IRC. 

\begin{figure}
\figurenum{9}
\epsscale{0.85}
\plotone{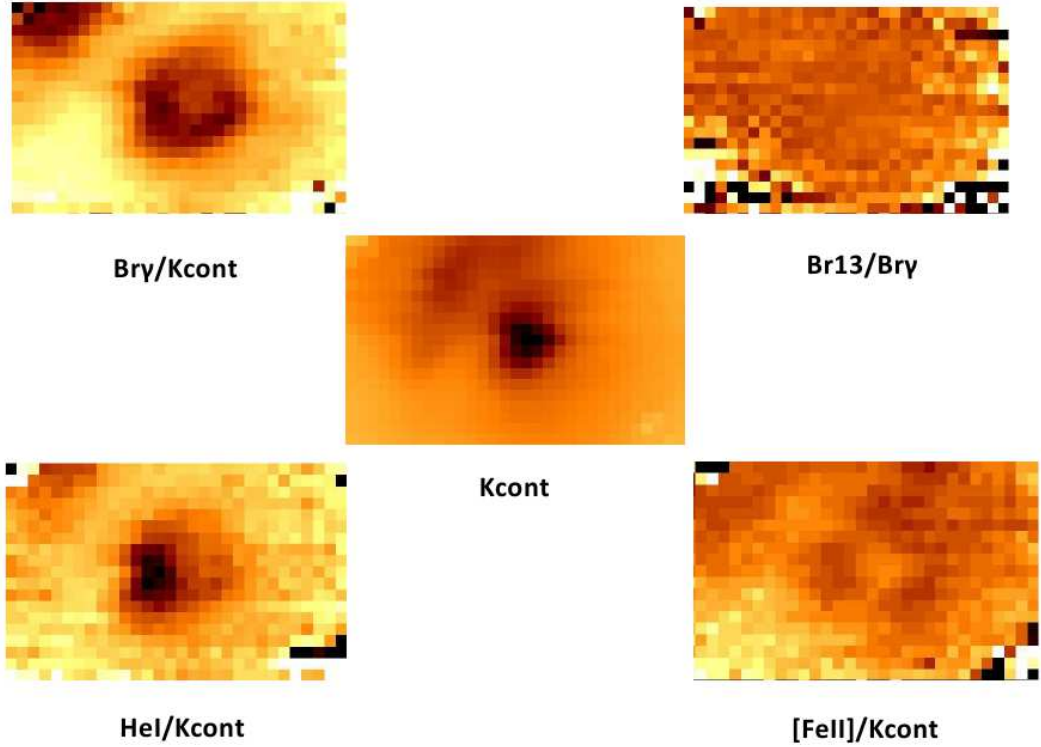}
\caption{Emission line indices. Each panel samples a $4.9 \times 3.2$ arcsec 
area, with north to the left and east at the top. 
The Kcont image from Figure 8 is shown in the middle panel. 
{\it Top left hand corner:} The ratio of Br$\gamma$ to Kcont. Br$\gamma$ 
emission has a more extended distribution on the sky than the stars that dominate the 
$K$ continuum, with the Br$\gamma$/Kcont ratio peaking along the western edge of the 
IRC, as well as in the north east corner of the IFU field. 
{\it Top right hand corner:} The ratio of Br13 to Br$\gamma$. 
There is a tendency for this ratio to decrease with increasing distance from 
the center of the IRC. {\it Lower left hand corner:} The ratio of HeI to Kcont. 
HeI emission has a more extended distribution than the light from 
the stars that dominate the NIR continuum near the IRC. The 
HeI/Kcont ratio peaks near the northern edge of the IRC, 
suggesting that this is an area where Wolf-Rayet stars 
are concentrated with respect to the stars that dominate Kcont. 
The area where the HeI/Kcont ratio is highest coincides with a peak in 
the Br$\gamma$/Kcont ratio. {\it Lower right hand corner:} The ratio of [FeII]$+$Br12 
to Kcont. Areas of residual [FeII] emission bracket the IRC. 
The excess [FeII] emission to the south east of the 
IRC is located along a ridge in the Br$\gamma$/Kcont ratio, although the 
Br$\gamma$/Kcont ratio there is lower than in the Br$\gamma$ arc to the west of 
the IRC.}
\end{figure}

	The Br$\gamma$ to Kcont ratio also peaks in the north 
east corner of the IFU field. The HeI/Kcont and [FeII]/Kcont ratios 
are also high in this area (see below). There are pockets with blue 
F128N--F164N colors in this area (Figure 1), and these may be areas of 
star-forming activity powering the Br$\gamma$ emission.

	The photons that excite Br$\gamma$ emission likely originate from 
O stars, and Kornei \& McCrady (2009) find that the IRC contains a large 
number of stars with ages $\sim 6$ Myr based on the strength of Br$\gamma$ emission. 
To the extent that Br$\gamma$ emission tracks young stars then the 
arc to the west of the IRC is where the ratio of hot young stars to the stars that 
dominate the NIR stellar continuum is locally highest. It is thus worth noting that 
the CO index in the western arc of Br$\gamma$ emission is not weaker than in the 
main body of the IRC -- the continuum from hot stellar photospheres evidently does 
not veil the CO features in the outer regions of the IRC. However, veiling is 
likely an issue in the main body of the IRC (Section 6).

\subsubsection{Br13/Br$\gamma$}

	The relative strengths of lines in the Brackett series can be used to 
investigate reddening, although there are caveats. 
While the Brackett lines sample wavelengths where 
the effects of extinction are greatly diminished when compared with visible 
wavelengths, there may be areas of extremely high extinction 
that are largely opaque in the NIR, and measuring the total line-of-sight 
extinction towards these areas will be problematic. Moreover, while wide 
wavelength coverage provides the greatest leverage for estimating extinction, 
the lines progressively weaken towards higher-order transitions, resulting in 
a trade-off between sensitivity to reddening variations and noise in 
line strength measurements. In addition, lines that are at different wavelengths 
may not originate at a common optical depth, skewing extinction 
estimates. Finally, relative line strengths depend 
on the electron temperature and density, and if these vary substantially 
across a region then the Brackett decrement will not be a faithful probe of 
extinction.

	The ratio of the Br13 and Br$\gamma$ indices is shown in the 
upper right hand corner of Figure 9. Br13 was selected for comparison 
with Br$\gamma$ by balancing the desire for a wide wavelength baseline against 
line strength. There are also well-defined continuum intervals on either side of Br13.

	While the Br13 to Br$\gamma$ ratio shows only subtle variations 
and there is considerable noise in this ratio outside of the IRC, aperture 
measurements reveal a systematic tendency for the Br13 to 
Br$\gamma$ ratio to decrease with increasing distance from the IRC. If 
the electron temperature and density are more-or-less constant across the area 
sampled then this behaviour suggests that extinction becomes lower 
with increasing distance from the IRC. This runs counter to the 
trend for color to become redder outside of the IRC (e.g. Figure 1). We speculate 
that the Brackett decrement near the IRC is tracking differences 
in ionization properties, rather than differences 
in line-of-sight extinction. In fact, the ratio of HeI line emission strength to Kcont 
-- discussed in the next section -- changes with distance from the center of IRC, 
indicating that very hot stars are not uniformly distributed near the IRC.

\subsubsection{HeI/Kcont}

	Prominent HeI $2.06\mu$m emission suggests that W-R 
stars are present in and around the IRC. The presence 
of these stars indicates that there has been large-scale 
star formation near the center of NGC 253 within the past 
few Myr. The ratio of the HeI and Kcont indices is shown in the lower 
left hand corner of Figure 9. The W-R stars that presumably power 
the HeI emission have a distribution that does not follow the 
stars that dominate Kcont. There is an area of localized high HeI 
emission with respect to Kcont to the north west of the IRC, and this is likely 
where the ratio of W-R stars to the stars that dominate the continuum near the IRC 
is highest. In addition to being an area of excess (when compared with stellar light) 
Br$\gamma$ and [FeII] emission (see below), the pocket of enhanced 
HeI emission is close to an area with blue F128N--F164N color in Figure 1, 
which is consistent with there being a concentration of hot, blue stars.

\subsubsection{[FeII]/Kcont}

	[FeII] emission is excited in shocks, 
and is usually associated with SNRs in star-forming environments. While the 
$1.64\mu$m [FeII] line is one of the strongest emission features in the NGC 
253 NIR spectrum, it is blended with Br12 in the GNIRS data. Based on the peak 
intensity of the [FeII]$+$Br12 blend, coupled with the relative strengths of Br11 and 
Br13, it appears that Br12 contributes $\sim 20 - 25\%$ of the total signal in 
the [FeII] index as defined in Table 1. This level of contamination will 
bias the [FeII] index to follow the distribution of Brackett hydrogen emission.

	The [FeII]/Kcont ratio is mapped in the 
lower right hand corner of Figure 9. As with the other emission indices, 
[FeII] does not follow the Kcont light. There are four areas where 
there are localized peaks in the [FeII/]/Kcont ratio. 
One is in the north east corner of the IFU field, where 
Br$\gamma$/Kcont and HeI/Kcont also peak. The other three areas surround the 
IRC. Two of these are along the western edge of the IRC and 
are close to areas that have a high Br$\gamma$/Kcont ratio, 
and one of these in turn coincides with the peak in the HeI/Kcont ratio. 
The fourth peak in the [FeII]/Kcont ratio 
is located to the south east of the IRC. While the 
[FeII]/Kcont ratios in the four regions are comparable, this is not 
the case for the Br$\gamma$/Kcont ratio, indicating that the [FeII]
index does not parrot Br$\gamma$ emission.

	The distributions of Br$\gamma$, HeI, and [FeII] emission are 
compared in Figure 10, where contours obtained from the Br$\gamma$/Kcont 
and HeI/Kcont images are superimposed on the [FeII]/Kcont image. 
The local peaks in the [FeII]/Kcont ratio around the IRC are offset slightly from 
those in Br$\gamma$/Kcont and HeI/Kcont, in the sense that the [FeII] emission 
peaks at slightly larger distances from the center of the IRC when compared with 
Br$\gamma$ or HeI. Such an offset could result if the [FeII] emission is tracking 
a shock front -- presumably powered by SNe activity -- that is propogating 
outwards from the star-forming regions. We note that Gunthardt et al. (2015) 
conclude that there is an outflow from the IRC based on the 
distribution of [NeII] $12.8\mu$m and Si $8.8\mu$m emission.

\begin{figure}
\figurenum{10}
\epsscale{0.75}
\plotone{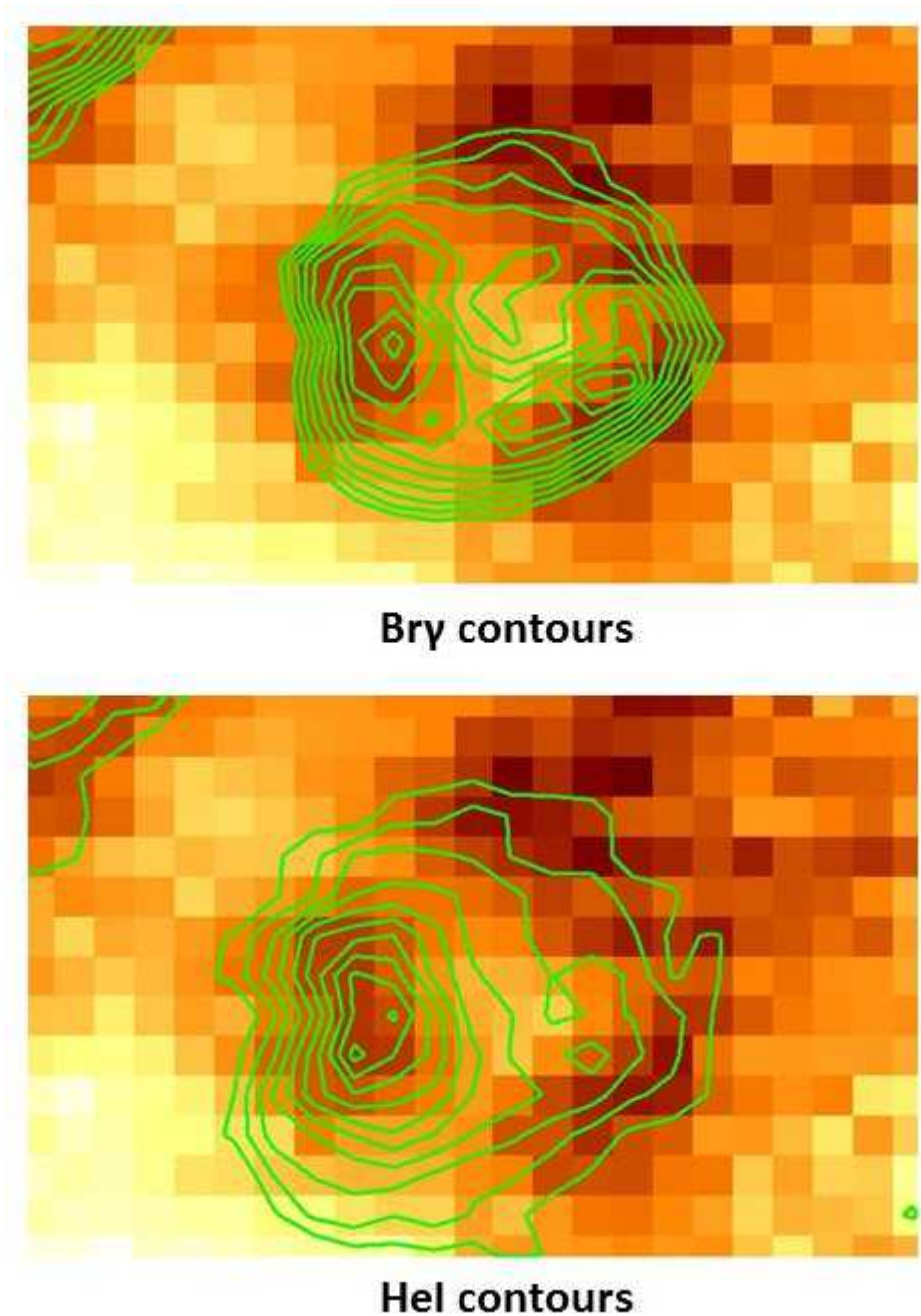}
\caption{Comparing the distributions of [FeII], Br$\gamma$, and HeI emission. 
The base image in each panel is a $4.1 \times 2.3$ arcsec sub-section of the 
the [FeII]/Kcont image from Figure 9. The green contours were obtained from the 
Br$\gamma$/Kcont (top panel) and HeI/Kcont (bottom panel) images in Figure 
9. Localized peaks in the [FeII] distribution near the IRC are offset to larger 
distances from the center of the IRC than the peaks in the Br$\gamma$ and HeI 
distributions.}
\end{figure}

\section{COMPARISONS WITH MODELS}

	Comparisons with model spectra can provide insights into the stellar content 
in and around the IRC. Highly evolved red stars are the dominant contributors to 
the photospheric NIR light from simple stellar populations (SSPs) 
that have ages $\geq 8 - 10$ Myr. Uncertainties in 
the physics associated with the most advanced stages of 
stellar evolution (e.g. review by Herwig 2005) thus play a significant role in 
interpreting the spectra of such systems. The well-known age-metallicity 
degeneracy adds additional uncertainties, as the color (and hence spectroscopic 
properties) of evolved stars depend on both age and metallicity. 

	Spectra constructed from SSP models described 
by Leitherer et al. (1999 -- hereafter STARBURST99) and Cordier 
et al. (2007 -- hereafter BaSTI) are compared with the observed spectra of 
NGC 253 and NGC 7793 in this section. The model spectra have a resolution 
$\lambda/\Delta\lambda \sim 200$, and so even prominent features such as the first 
and second-overtone CO bands are blended with other molecular and/or 
atomic transitions. The baseline models assume solar metallicities 
and constant SFRs (cSFRs) from the present day to a time in the 
past. Given that the NGC 253 and NGC 7793 spectra were not flux calibrated, the 
comparisons focus on matching the absorption characteristics in continuum-corrected 
spectra, and no effort is made to estimate a SFR from the models. In any event, 
an estimate of the SFR depends on the star-forming history 
(SFH), which is not well-constrained by the NIR 
spectrum. The models are also used to examine the effect of veiling from very young 
populations, as this can have a significant impact on the depth of absorption 
features. 

\subsection{The Models}

	cSFR models constructed from STARBURST99 SSP spectra 
are shown in the left hand panels of Figures 11 ($H-$band) and 12 ($K-$band). The 
models have had the continuum removed, and have been shifted vertically for 
display purposes. The models are based on the solar metallicity 
Padova isochrones described by Bertelli et al. (1994), with TP-AGB evolution 
implemented using the procedures described by Vazquez \& Leitherer (2005). 
A two-component Kroupa (2001) mass function is adopted.

\begin{figure}
\figurenum{11}
\epsscale{1.00}
\plotone{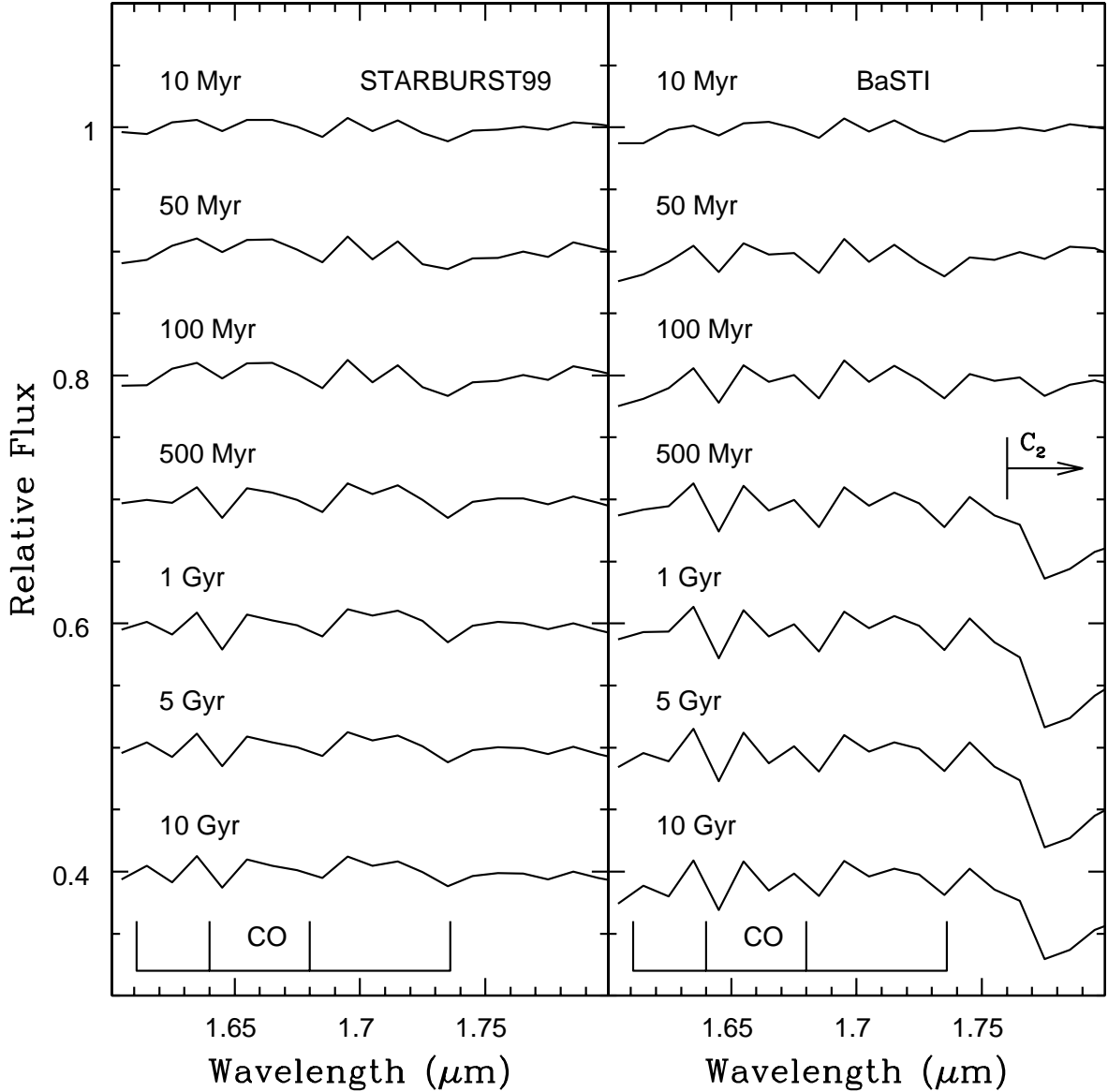}
\caption{Solar metallicity cSFR models constructed from STARBURST99 
and BaSTI SSP $H-$band spectra. The models have had the continuum removed, and 
have a resolution $\lambda/\Delta\lambda \sim 170$. 
The locations of CO band heads are indicated. The 
models assume a continuous SFR (cSFR) from the present day to the age indicated; 
for example, the 1 Gyr model reproduces the integrated spectrum of a system that 
evolves with a cSFR from 1 Gyr ago up to the present day. 
The 10 Myr STARBURST99 model is repeated in the right hand panel, as the BaSTI 
models do not track evolution younger than 30 Myr. Changes in age cause variations 
in the strengths of most features at the few percent level. 
C$_2$ absorption with a depth of a few percent is seen  
at wavelengths $> 1.77\mu$m in BaSTI models with ages $\geq 500$ Myr.}
\end{figure}

\begin{figure}
\figurenum{12}
\epsscale{1.00}
\plotone{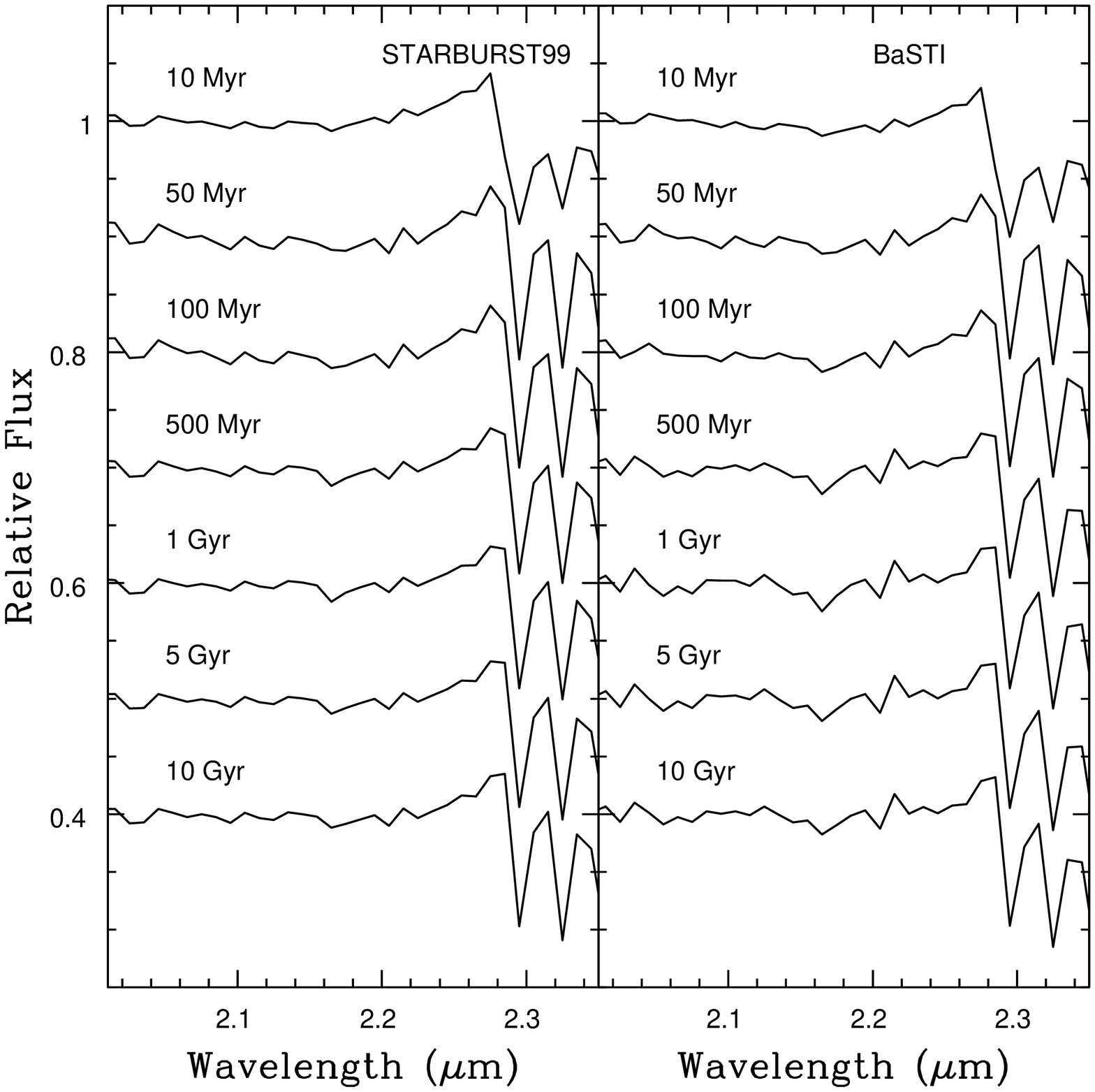}
\caption{Same as Figure 11, but showing $K-$band models. 
The first-overtone CO bands are the most prominent absorption features in this 
wavelength range. The depths of the CO band heads differ by a few percent with age, 
and are deepest in the 50 and 100 Myr models. The strengths of many 
features between 2.0 and 2.25$\mu$m, which are 
heavily blended molecular and atomic transitions, also vary with age.}
\end{figure}

	cSFR models constructed from SSP spectra in the BaSTI 
compilation are shown in the right hand panels of Figures 11 and 12. 
The model spectra are based on the solar metallicity 
isochrones described by Pietrinferni et al. (2004), and include TP-AGB evolution as 
described by Cordier et al. (2007). A Chabrier (2001) mass function is 
adopted. The BaSTI models do not include systems with ages $< 30$ 
Myr, and so the STARBURST99 SSP models in that 
age range were used when constructing the BaSTI cSFR models.

	The first and second-overtone CO bands are the dominant features in the 
models. These are deepest in the 50 and 100 Myr spectra, reflecting the large 
contribution to NIR light made by luminous low surface-gravity RSGs at these ages. 
The differences in the CO band strengths due to age are not large, and the 
first-overtone CO bands are only a few percent deeper 
in the 50 and 100 Myr models when compared with models in which star 
formation continues over a longer time frame. The 
depths of the CO bands do not change significantly in the cSFR models with ages 
$\geq 1$ Gyr, and so the CO bands contain little information about the SFH of older 
composite systems. That these features are not sensitive to the SFH makes them of 
potential interest as metallicity indicators. 

	There are age-related differences in the strengths of weaker 
absorption features in both spectral regions. 
These weaker features are blends of molecular and atomic transitions 
and their depths vary by a few percent for ages $\geq 30$ Myr. The peak between 
the (2,0) and (3,1) CO bands is also noticeably smaller in the 10 Myr model 
than in the other models.

	Significant differences between the STARBURST99 and BaSTI 
models can be seen, with the BaSTI models having 
larger age-related variations and deeper absorption features. 
These differences are greatest in the $H-$band, and 
the most striking difference is associated with the Ballick-Ramsey 
C$_2$ band. This feature is abscent in the STARBURST99 models, 
but is clearly present in the BaSTI cSFR models with ages $> 500$ Myr. While C stars 
are usually associated with systems that have sub-solar 
metallicities, moderate numbers of C stars might be expected
in solar metallicity systems. Karakas (2014) finds that solar metallicity 
stars with masses between 2 and 4.5M$_{\odot}$ will become C stars. 
The ratio of the number of C stars to the number of O-rich AGB stars of spectral-type 
M -- C/M -- in intermediate age solar-metallicity systems is 
$\sim 0.07$ in her models. Studies of C stars in the Galactic disk suggest that the 
lower mass limit for C star formation may even extend to lower values than predicted 
by the Karakas models (e.g. Feast, Whitelock, \& Menzies 2006), potentially 
increasing the possible constribution from C stars to the integrated NIR light. 

	The BaSTI cSFR models predict that C$_2$ absorption should be present if 
star-forming activity has occured over a time interval 
of at least a few hundred Myr. These models further suggest 
that an exceptionally high SFR during intermediate epochs is not required to 
produce detectable C$_2$ absorption. Rather, C$_2$ might be 
detectable if the SFR during intermediate epochs is simply 
comparable to that at other epochs and the noise level in the spectrum 
is no larger than a few percent -- if the BaSTI models are correct then C$_2$ 
absorption with a depth of a few percent should be common in the spectrum of solar and 
sub-solar metallicity systems that have experienced extended 
periods of star formation during intermediate epochs. That C$_2$ absorption 
is present in the model spectra of systems that have experienced a cSFR over a large 
fraction of the Hubble time is probably due to 
C stars being intrinsically brighter in the NIR than older TP-AGB stars. 
C stars then contribute significantly to the NIR light of composite systems even 
though they form only over a limited range of stellar masses. 

\subsection{Comparisons with Observed Spectra}

	The NGC 253 East and West spectra differ in the strengths of some emission 
features, but have very similar absorption features. The means of the NGC 253 East 
and West spectra were thus computed, and the results will be referred to 
as the circumIRC (cIRC) spectra. The spectra of the IRC, the cIRC, and the nucleus of 
NGC 7793 are compared with various models in Figures 13 ($H$) and 14 ($K$). 
The GNIRS spectra have been smoothed to match the $0.01\mu$m resolution of the models.

\begin{figure}
\figurenum{13}
\epsscale{1.00}
\plotone{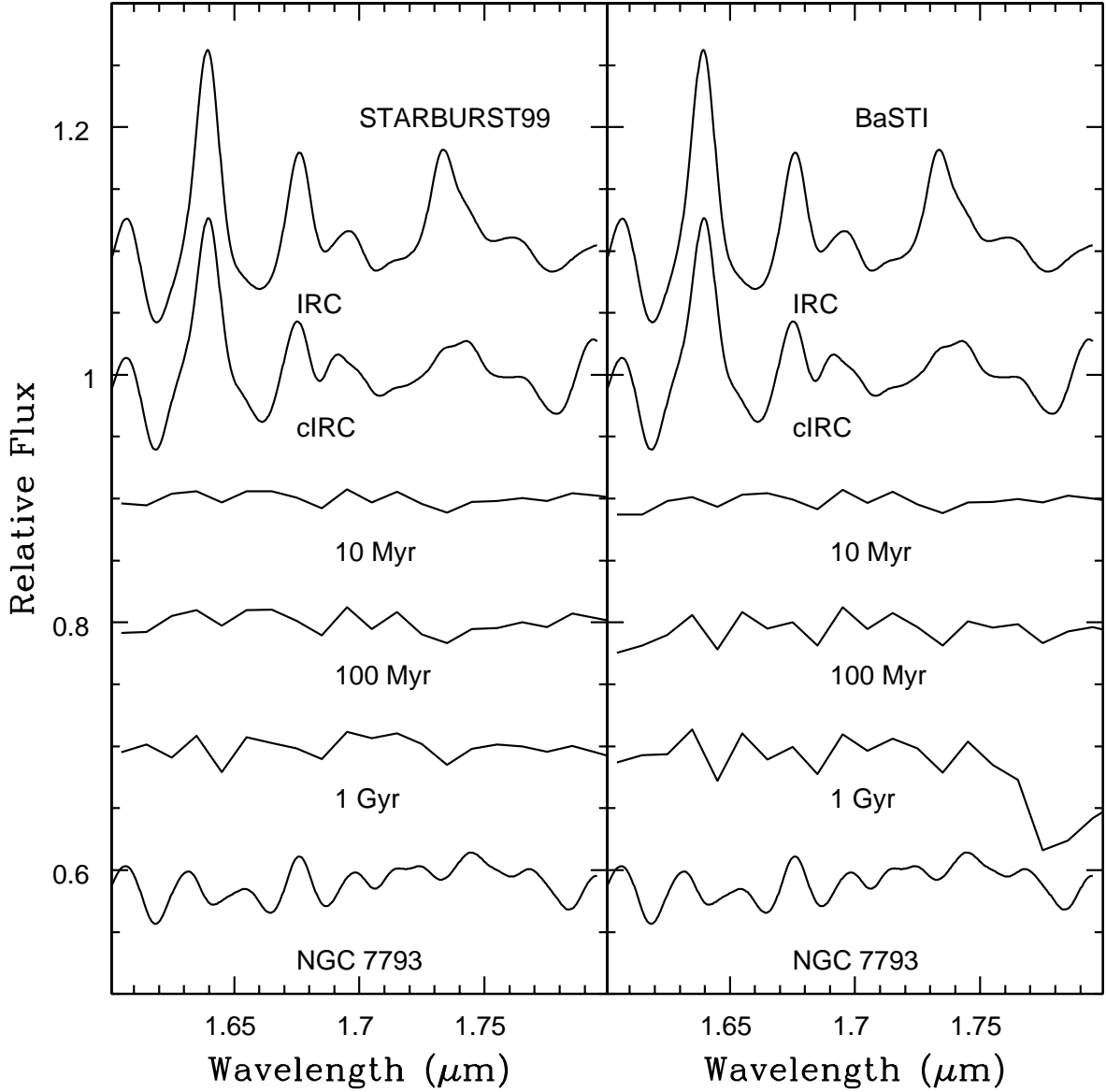}
\caption{Comparisons with $H-$band models. The GNIRS spectra 
have been smoothed to match the resolution of the models. While strong 
emission lines complicate comparisons between the NGC 253 spectra and the models, 
this is not a significant factor for NGC 7793, although weak emission lines may become 
apparent after smoothing (see text). Features in the NGC 7793 spectrum 
tend to be stronger than predicted by the solar metallicity STARBURST99 models.
There is an absorption feature at $\lambda > 1.78\mu$m in the NGC 7793 spectrum 
that coincides with the C$_2$ feature in the BaSTI models, but has a smaller 
amplitude. A similar absorption feature is also seen at this wavelength in the 
NGC 253 spectra.}
\end{figure}

\begin{figure}
\figurenum{14}
\epsscale{1.00}
\plotone{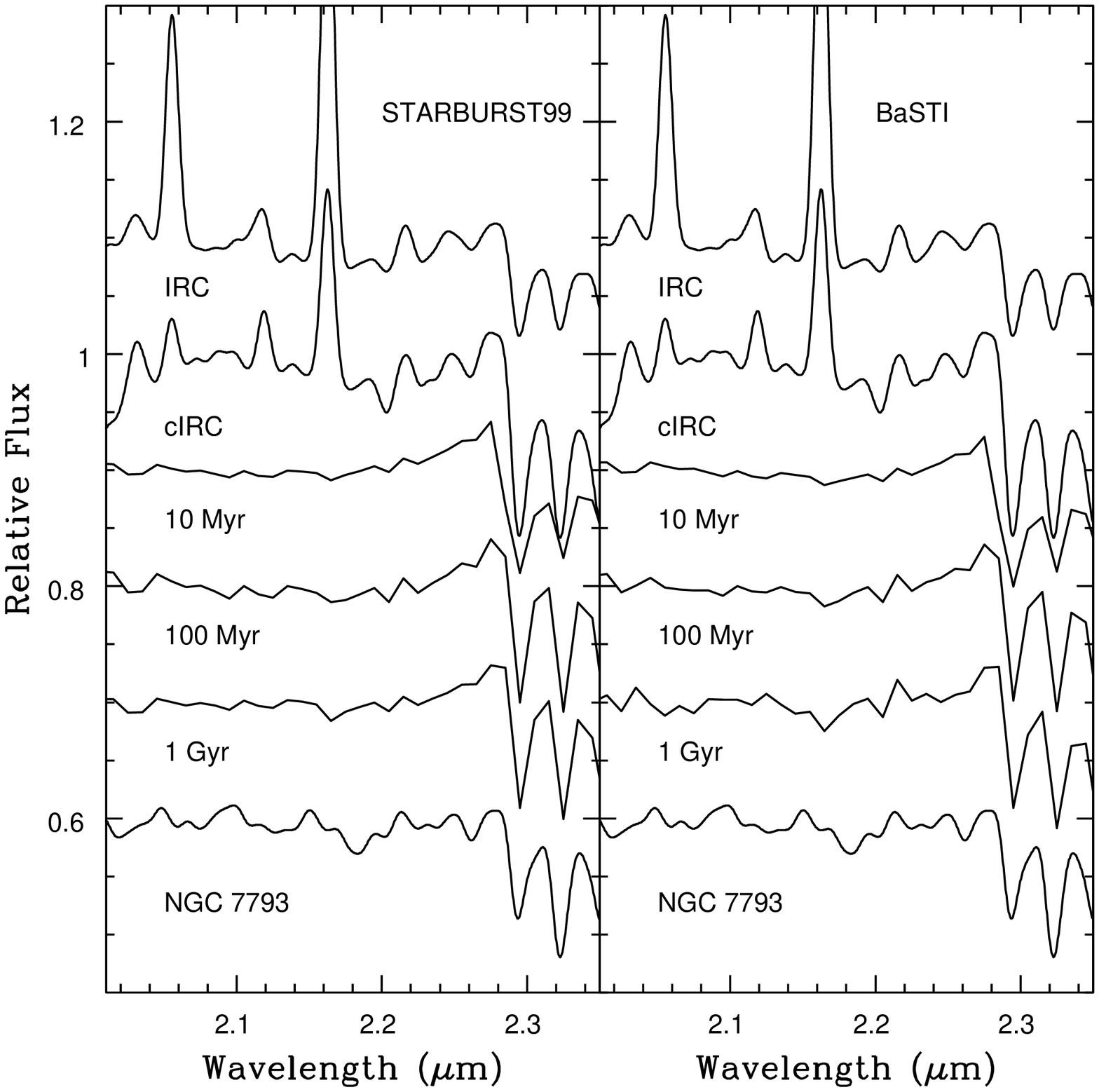}
\caption{Same as Figure 13, but showing $K-$band models. The 
first-overtone CO bands in NGC 7793 and the IRC are 
slightly weaker than predicted by the solar metallicity models. 
As noted in Section 6.2.2, veiling by hot stellar photospheric continua may 
weaken the depths of absorption features in the IRC spectrum by almost 10\%.
In contrast, the CO bands in the cIRC spectrum are deeper than 
in the 100 Myr and 1 Gyr models.}
\end{figure}

\subsubsection{NGC 7793}

	The NGC 7793 spectrum is considered first, as it is free of obvious emission 
features. There is mixed success in matching the NGC 7793 spectrum 
with the models. The absorption feature in the NGC 7793 spectrum near 
$1.76\mu$m overlaps with the C$_2$ feature in the BaSTI model spectra. 
However, many of the weaker features in the NGC 7793 spectra do not have 
obvious counterparts in the models. This might be attributed in part to 
binning, as the method by which binning is done can alter the appearance of 
low resolution spectra (e.g. Figure 6 of Davidge et al. 2015). Binning 
will presumably be less of a factor for very strong features, such as 
the depths of the first-overtone CO bands, and the CO bands in the NGC 7793 
spectrum are slightly weaker than in the solar metallicity models. 

	That the $K-$band model spectra have deeper CO bands than in the NGC 
7793 nuclear spectrum might suggest a slightly sub-solar metallicity for the nucleus 
of NGC 7793. This being said, Bibby \& Crowther (2010) find a central O abundance 
of $12 +$log(O/H)$=8.6$ for NGC 7793, which is consistent with solar (e.g. Asplund 
et al. 2004). However, [O/Fe] has not yet been measured in NGC 7793. In addition, if 
the NIR light from the nucleus is dominated by stars that formed during intermediate 
epochs then the nucleus might have a lower luminosity-weighted metallicity than 
the present-day gas.

	Weak line emission may become evident when the NGC 7793 spectra are 
smoothed, as this increases the S/N ratio. In fact, subtle features appear in 
the smoothed NGC 7793 nuclear spectrum that coincide with emission features in 
the NGC 253 spectra. A modest population of hot young stars that power this emission
may then be present in the NGC 7793 nucleus. That only a modest number of young 
or intermediate age stars are present near the center of 
NGC 7793 is perhaps not surprising. The frequency of blue nuclei 
among nearby disk galaxies suggests that the duty cycle for central star formation 
is on the order of $\sim 1$ Gyr (e.g. Davidge \& Courteau 2002; Ho et al. 1997), 
and the weak emission line contribution might then indicate that the nucleus of 
NGC 7793 is currently between large-scale star-forming episodes. 

\subsubsection{NGC 253}

	The BaSTI and STARBURST99 models both underestimate the total 
amplitude of the (2,0) CO band head in the cIRC spectrum. 
The comparisons in Figure 14 thus suggest that 
the brightest RSGs in the cIRC have a metallicity that is slightly super-solar. 
The situation is different for the IRC, as the first overtone 
CO bands in that system are weaker than in the solar metallicity models. Given that 
the overall depths of the CO bands do not change greatly with SFH (Figure 12), and 
that veiling by nebular emission does not change the depths of absorption 
features by huge amounts (Section 4), then the difference in the CO 
depths between the models and the IRC spectrum might suggest that 
the IRC has a sub-solar metallicity. However, such a 
conclusion assumes that the NIR light is dominated by red stars.

	Veiling of the CO bands could be significant if a large fraction of 
the NIR light originates from hot young stars, which have a near-featureless NIR 
spectrum. Previous NIR spectroscopic studies of the central regions of 
NGC 253 have concluded that it is an area of intense recent star formation 
(e.g. Engelbracht et al. 1998). The presence of strong HeI and Br$\gamma$ emission 
throughout the central regions of NGC 253 indicates 
that hot, young stars are present in large numbers, and 
the areas of the IRC that have blue NIR colors in Figure 1 
indicate that hot stars contribute significantly to the NIR light there. 

	The fraction of the mass of recently formed stars 
that can veil the absorption spectrum depends on the assumed age of 
the IRC. Consider the effect of adding a population with an age of 4 
Myr (i.e. a population with massive stars that might power HeI emission) to that of 
the cSFR STARBURST99 models with ages of 50 Myr, 500 Myr, and 10 Gyr. 
These ages were selected to represent different evolutionary cases 
for NGC 253. The 50 Myr age explores the hypothesis that the IRC is 
relatively young, as might be the case if the starburst was only recently triggered. 
The 500 Myr model is included to represent the case in which the IRC formed as 
part of a mature starburst that has been continuing for a few disk crossing times. 
Finally, the 10 Gyr model explores the case in which the IRC is an old structure -- 
possibly the nucleus of NGC 253 -- that has experienced periodic levels of elevated 
star-forming activity throughout the age of the Universe, and is once again the 
site of renewed star-forming activity. If the absorption 
feature in the IRC spectrum near $1.78\mu$m is due to C$_2$ then this favors the 
500 Myr and 10 Gyr scenarios.

	Assuming that the red stars in the IRC have solar metallicities, then 
the fractional contributions that a 4 Myr population must make to the total 
mass of the IRC to diminish the depth of the (2,0) CO band in the solar 
metallicity model spectrum to that seen in the IRC are 70\% (50 Myr assumed 
age for the IRC), 30\% (500 Myr assumed age), and 10\% (10 Gyr assumed age). The 
fractional contribution from the 4 Myr population drops with increasing IRC age 
because the M/L ratio in $K$ grows with increasing age, so that a smaller mass of 
4 Myr stars is required to alter the spectrum of the underlying system 
as its age increases. Still, the assumed age of the IRC notwithstanding, the IRC is 
evidently being observed at a special time when it is experiencing rapid growth in 
stellar mass. If (1) the current level of star formation has remained constant 
during recent times and does not change over the near future, and 
(2) the stars that form in the IRC stay bound to that structure, then 
the times required to double the original stellar mass of the IRC 
are 3 Myr (50 Myr IRC age), 9 Myr (500 Myr age), and 36 Myr (10 Gyr age). 

	When compared with the IRC, there is smaller fractional contamination by hot 
stars in the cIRC spectrum, and so the absorption features are less affected by 
veiling. The HeI line in the cIRC spectrum is only one-third as strong as in the IRC 
spectrum, while the Br$\gamma$ line is roughly half as strong as in the IRC. If the 
light from the 4 Myr component in the models discussed above is decreased by a factor 
of 3 then the (2,0) CO band strength is diminished by only $\sim 3\%$, as opposed 
to the 7\% change that is predicted for the IRC.

	If the cIRC spectra sample a larger mix of stars with ages $\geq 8$ Myr than 
the IRC spectra then a larger fraction of the NIR light in the cIRC will come from 
bright RSGs. A larger fractional contribution from very young RSGs -- 
with a resulting smaller fractional contribution from hot stars -- 
could explain (1) why the CO bands in the East and West spectra 
have similar depths despite differences in the 
strength of Br$\gamma$ emission, and (2) why the CO bands in the pockets of strong 
HeI and Br$\gamma$ emission seen in Figures 8 and 9 are not weaker than in the 
surrounding areas. Based on NIR colors, Engelbracht et al. (1998) argue that there is 
no significant veiling of the spectrum of the center of NGC 253 by a featureless 
continuum, and this is more-or-less consistent with the subdued hot stellar continuum 
level inferred here for the cIRC region. In contrast, the F128N--F164N color data 
in Figure 1 clearly shows that the IRC contains regions with blue NIR colors, where 
veiling is expected to be significant.

	We close this section by noting that the addition of a hot stellar 
continuum will weaken other absorption features, including the trough at 1.78$\mu$m 
that might belong to C$_2$. Adding a hot stellar continuum to the 
cIRC $H-$band spectrum so as to reproduce the depths of the first 
overtone CO bands in the IRC will weaken C$_2$ absorption by 50\%. This 
is more-or-less consistent with the relative depths of the candidate C$_2$ features in 
the cIRC and IRC spectra in Figure 13. Veiling of this size will also bring 
the depth of C$_2$ absorption in the BaSTI models into better agreement with 
what is seen in the cIRC and IRC spectra.

\section{DISCUSSION \& SUMMARY}

	The stellar content near the center of the starburst 
galaxy NGC 253 has been investigated using NIR IFU spectra recorded with GNIRS on GS 
during sub-arcsec seeing conditions. The GNIRS IFU samples a $\sim 120$ arcsec$^2$ 
area, which corresponds to $\sim 0.03$ kpc$^2$ at the distance of NGC 253. 
The IRC, which defines the photometric center of the galaxy, and several 
nearby sub-structures are sampled.

	The analysis of integrated NIR light is at present one of the few 
means of examining the stellar content of the crowded central regions of 
NGC 253. The use of an IFU allows the spectrum of the IRC to be de-coupled from 
that of the surroundings, and the spatial distribution of stellar types to 
be mapped. It is clear that the IRC is not an SSP. As noted by Kornei \& McCrady 
(2009), the presence of deep CO absorption bands are indicative of RSGs, which 
form in systems that have an age of at least 8 -- 10 Myr. However, the 
presence of Br$\gamma$ and HeI emission indicate that IRC has been the site 
of highly elevated levels of star formation within the past $\sim 10$ Myr, 
The IRC thus contains stars with an age spread of at least a few Myr. Barring a 
termination of star-forming activity then the age spread will increase with time. 
Indeed, the total stellar mass of the IRC is expected to double over time scales of 
a few Myr to a few tens of Myr if the pace of star formation that occured during the 
past 10 Myr continues unabated. 

	The observations presented here indicate that the youngest stars are not 
well-mixed with the stars that contribute to the NIR continuum in and around the IRC. 
Excess Br$\gamma$ emission with respect to light from the stars that 
dominate the NIR continuum is found in an arc along the western boundary of the 
IRC, while concentrated HeI emission is seen to the north of the IRC. 
Recent episodes of star formation have thus not been restricted to the 
center of the IRC, but have also occured in its outer regions. 

	A key issue is the age of the IRC. Is it a young -- probably transient -- 
stellar ensemble, or is it a much older structure that may be the 
nucleus of NGC 253? In Section 7.1 we discuss the recent SFH of the IRC, 
and consider its fate if it is a young structure. In Section 7.2 
the possibility that the IRC is the old nucleus of NGC 253, 
as suggested by Gunthardt et al. (2015), is considered. 

\subsection{The Recent SFH of the IRC}

	That recent star formation has occured at different locations 
around the IRC suggests that it is likely not a `cluster', but 
rather a compact star-forming complex containing 
distinct sub-structures. Only the sub-structures in the outer region of the 
system are detected with GNIRS, due to crowding in the denser inner regions. 
Sub-structuring on finer scales is seen in the WFC3 images in Figure 1, which 
have angular resolutions that are $\sim 3 - 4\times$ finer than the 
GNIRS data.

	An age dispersion of at least a few Myr is expected in young systems. 
There is evidence for periods of star formation in young Local Group clusters 
that can last up to at least a few Myr. One of the largest young systems in the 
Galaxy is Westerlund 1, and it has an age dispersion of a few 
Myr (Kudryavtseva et al. 2012; Lim et al. 2013). A similar age spread is seen 
in Westerlund 3 (Bik et al. 2014). Such a short time scale for the formation of a 
second stellar generation in large star clusters is consistent with simulations of 
thermally stable clumps that can cool and form stars as a result of self-shielding 
(Palous et al. 2014). 

	Star formation may extend over much longer timescales than is 
seen in young Galactic clusters. The largest nearby 
star-forming complex to be surveyed in depth is 30 Doradus, and de Marchi et al. 
(2011) find an age dispersion of $\leq$ a few tens of Myr. Signatures of extended 
star formation are found in the spectra of some extragalactic clusters (e.g. 
Peacock et al. 2013, but see also Bastian et al. 2013). 
Conroy \& Spergel (2011) argue that star clusters with masses 
$> 10^4$ M$_{\odot}$ can retain star-forming material for extended periods of 
time, although the escape velocity prior to the onset of mass segregation 
may provide a better predictor of gas retention than total cluster mass 
(Goudfrooij et al. 2014). Indeed, while there is photometric evidence for a range of 
stellar ages in some massive LMC clusters (e.g. Keller et al. 2012; Li et al. 2014), 
there are exceptions (e.g. Bastian \& Silva-Villa 2013). 
Multiple episodes of star formation have also been suggested as a means of explaining 
the chemical properties of stars in globular clusters, and CMDs indicate that 
many globular clusters are not SSPs (e.g. review by Gratton et al. 2012). 

	If the IRC has a young age then it may not be long-lived. 
Fall \& Chandar (2012) present evidence that 
the majority of clusters are disrupted very early in their lives, and the 
tidal environment near the center of NGC 253 may not be conducive to the 
long term survival of even a massive cluster. It is tempting to compare the 
IRC with the Arches cluster, which has an age of only a few Myr but shows signs of 
tidal disruption in its global mass function (Stolte et al. 2005) and radial structure 
(Habibi et al. 2013). However, the Arches is roughly an order 
of magnitude less massive than the IRC. 

	That only modest levels of structuring are seen in the cIRC region does 
not indicate that the young stars found there at the present day formed {\it in 
situ} as part of a diffusely distributed field population. Rather, these young stars 
more likely formed in clusters that are now either defunct or largely disrupted. Stars 
that form in the outer regions of the IRC may be susceptible to being calved off, 
and so these areas may also be a source of future stellar generations in the cIRC 
environment.

\subsection{Is the IRC Old?}

	Gunthardt et al. (2015) argue that the IRC is the nucleus of NGC 253. 
Indeed, the integrated brightness and angular size of the IRC 
are similar to those of other galactic nuclei. A complicated SFH 
also does not preclude the IRC from being the nucleus, as it 
is not uncommon for the nuclei of late-type galaxies to host 
star clusters (e.g. Boker et al. 2002; Georgiev \& Boker 2014). 
This being said, there is a tendency for the photometric sizes of the nuclei of 
late-type spirals to vary with wavelength, in the sense that 
most systems tend to have smaller sizes towards shorter wavelengths, suggesting 
a tendency for younger stars to be more centrally concentrated than the main 
stellar body (e.g. Georgiev \& Boker 2014). This is not the case in the IRC.

	If the IRC is the nucleus of NGC 253 
then a substantial age spread would be expected. Deep CN and C$_2$ bands 
are NIR spectroscopic signatures of an underlying intermediate age population 
that would indicate that the IRC has survived disruption and formed 
stars for at least many hundreds of Myr. It is thus interesting that there is an 
absorption feature in the IRC spectrum near 1.78$\mu$m that may be due to C$_2$. 
In Section 4 it is argued that this absorption is likely 
not due to DIBs, and this could be checked using higher spectral-resolution 
observations of the center of NGC 253 in the 1.7 -- $1.8\mu$m wavelength interval. 

	The depths of the CO bands in the cIRC spectrum are consistent with a 
near-solar metallicity, and so the detection of C stars in significant 
numbers near the center of NGC 253 might be surprising. However,
models of C star evolution investigated by Karakas (2014) indicate that C stars 
can form at solar and super-solar metallicities, although the results for 
super-solar metallicities depend critically on the Helium content. The 
Karakas (2014) models predict that C stars in solar metallicity systems 
have progenitor masses between 2 and 4.5 M$_{\odot}$, which corresponds to 
main sequence turn-off ages $\sim 0.2 - 1.8$ Gyr. Based on the relative 
lifespans of C stars when compared with the entire AGB phase, then 
the ratio of C stars to M giants predicted by Karakas (2014) for a solar 
metallicity system in this age range is $\sim 0.07$. If the M giants and C stars 
in such a system have comparable intrinsic brightnesses then a modest 
C$_2$ feature would be expected in the integrated NIR spectrum.

	Davidge (2010) used NIR colors to identify C star candidates throughout 
the disk and extraplanar regions of NGC 253, and find that C stars may be present 
in large numbers throughout the galaxy. Near the tip of the AGB the C star 
candidates contribute $\sim 20\%$ of the $K-$band light at galactocentric radii 
6 -- 10 kpc. However, a problem when identifying C stars from broad-band photometry 
alone is that the threshold color that separates C stars from M giants is a function 
of metallicity (e.g. discussion in Boyer et al. 2013), and so metal-rich M giants 
may have colors that overlap with those of more metal-poor C stars. The C star 
contribution to the total light estimated from photometric surveys alone 
is thus an upper limit. MIR photometry may provide a means 
of distinguishing between very cool M giants and C stars, as the SED of 
the most evolved stars in the $3 - 8\mu$m wavelength interval is sensitive 
to the mineralogy of the circumstellar envelope (Dell'Agli et al. 2014).

	In Section 6 it was demonstrated that hot young stars likely contribute 
significantly to the $K-$band light from the IRC. The models used in Section 6 
involve only two components -- a very young SSP coupled with an idealised 
system that has experienced a cSFR. Still, these models suggest that if the IRC is the 
nucleus of NGC 253, and thus contains stars that formed 
over a large fraction of the age of the Universe, then 
very young stars account for 10\% of its stellar mass. 
This opens the possibility that the nucleus is currently 
experiencing an episode of very rapid growth in terms of stellar mass. If the 
nucleus contains a massive black hole then the present day might also be a time during 
which the black hole is also accreting substantial mass. While Gunthardt et al. 
(2015) find no signs of nuclear activity, the orientation of NGC 253 on the sky 
may mean that any acccretion activity is hidden.

\acknowledgements{It is a pleasure to thank the anonymous referee for providing 
a prompt and comprehensive report that greatly improved the paper.}

\end{document}